\documentclass[a4paper,fleqn]{cas-sc}
\usepackage{savesym}
\savesymbol{fax}
\savesymbol{Hermaphrodite}
\usepackage[authoryear,longnamesfirst]{natbib}
\usepackage[T1]{fontenc}
\usepackage{lscape}
\usepackage{algorithm2e}
\usepackage{enumitem}
\usepackage{amsmath, bm}
\usepackage[utf8]{inputenc}
\usepackage[rightcaption]{sidecap}
\usepackage{times}
\usepackage{helvet}
\usepackage{courier}
\usepackage{listings}
\usepackage{amsmath}
\usepackage{amsfonts}
\usepackage{graphicx}
\usepackage{hhline}
\usepackage{enumerate}
\usepackage{multirow}
\usepackage{array}
\usepackage{color}
\usepackage{float}
\usepackage{wrapfig}
\usepackage{nopageno}
\DeclareMathOperator{\diag}{diag}
\usepackage{subcaption}
\usepackage{pifont}
\newcommand{\cmark}{\ding{51}}%
\newcommand{\xmark}{\ding{55}}%
\usepackage{caption}
\usepackage{algpseudocode}

\usepackage[show]{chato-notes}
\usepackage{bm} 

\newcommand{\mymtx}[1]{\bm{\mathrm{#1}}}
\newcommand{\set}[1]{\mathcal{#1}}
\newcommand{\wang}[1]{\textcolor{black}{#1}}
\newcommand{\siwei}[1]{\textcolor{black}{#1}}
\newcommand{\iadh}[1]{\textcolor{black}{#1}}
\newcommand{\craig}[1]{\textcolor{black}{#1}}
\newcommand{\liu}[1]{\textcolor{black}{#1}}

\newcommand{\iadhr}[1]{\textcolor{black}{#1}}

\newcommand{\siweir}[1]{\textcolor{black}{#1}}
\newcommand{\craigr}[1]{\textcolor{black}{#1}}
\newcommand{\xw}[1]{\textcolor{black}{#1}}
\newcommand{\io}[1]{\textcolor{black}{#1}}
\newcommand{\swr}[1]{\textcolor{black}{#1}}

\newcommand{\swai}[1]{\textcolor{black}{#1}}
\newcommand{\craigai}[1]{\textcolor{black}{#1}}
\newcommand{\xwai}[1]{\textcolor{black}{#1}}
\newcommand{\ioai}[1]{\textcolor{black}{#1}}
\newcommand{\swext}[1]{\textcolor{black}{#1}}
\newcommand{\xiw}[1]{\textcolor{black}{#1}}
\newcolumntype{P}[1]{>{\centering\arraybackslash}p{#1}}

\newcommand{\ipms}[1]{\textcolor{black}{#1}}

\newcommand{\blue}[1]{\textcolor{black}{#1}}
\newcommand{\yab}[1]{\textcolor{black}{#1}}

\newcommand{\ipmm}[1]{\textcolor{black}{#1}}

\begin{document}
\let\WriteBookmarks\relax
\def\floatpagepagefraction{1}
\def\textpagefraction{.001}
\shortauthors{Siwei Liu, Xi Wang, Craig Macdonald, Iadh Ounis}
\title [mode = title]{A Social-aware Gaussian Pre-trained Model for Effective Cold-start Recommendation}   

\author[1]{Siwei Liu}[orcid=0000-0002-7326-2883]
\ead{siwei.liu@mbzuai.ac.ae}

\author[2]{Xi Wang}[orcid=0000-0001-5936-9919]
\ead{xi-wang@ucl.ac.uk}

\author[3]{Craig Macdonald}[orcid=0000-0003-3143-279X]
\ead{craig.macdonald@glasgow.ac.uk}

\author[3]{Iadh Ounis}[orcid=0000-0003-4701-3223]
\ead{iadh.ounis@glasgow.ac.uk}

\affiliation[1]{organization={Mohamed bin Zayed University of Artificial Intelligence
, Department of Machine Learning},
    addressline={Masdar}, 
    city={Adu Dhabi},
    postcode={44737}, 
    country={United Arab Emirates}}  
\affiliation[2]{organization={University College London, School of Computer Science},
    addressline={Gower Street}, 
    city={London},
    postcode={WC1E 6BT}, 
    country={United Kingdom}} 
\affiliation[3]{organization={University of Glasgow, School of Computing Science},
    addressline={Lilybank Gardens}, 
    city={Glasgow},
    postcode={G12 8QQ}, 
    country={United Kingdom}}
    
\shorttitle{A Social-aware Gaussian Pre-trained Model for Effective Cold-start Recommendation}

\begin{abstract}
The \iadh{use of} pre-training is an emerging technique to enhance a \iadh{neural} model's performance, which has been \iadh{shown} to be effective for many \iadh{neural} language models such as \iadh{BERT}. This technique has also been \io{used} to \iadh{enhance the performance of recommender systems. }\siweir{In such recommender systems, pre-training models are used to learn a better initialisation for \io{both} users and items. }
However, recent \iadh{existing} \io{pre-trained recommender systems} tend to \iadh{only} incorporate the \iadh{user} interaction data at the pre-training stage, making it \iadh{difficult} to deliver good recommendations, especially when the interaction data is sparse. To alleviate this \iadhr{common} data sparsity issue, we propose to \siweir{\textit{pre-train}} the \iadh{recommendation} model \iadh{not only} with the interaction data \iadh{but also with other available information such as} the social relations \iadh{among users}, \iadh{thereby providing} the recommender \iadhr{system} \iadh{with} a better initialisation compared with solely relying on the \iadhr{user} interaction data. \iadh{We propose} a novel recommendation model, the \textit{Social-aware \siweir{Gaussian} Pre-trained} model (SGP), which encodes the \iadh{user} social relations and interaction data at the \textit{pre-training} stage in a Graph Neural Network (GNN). {Afterwards, in the subsequent fine-tuning stage, our SGP model adopts a Gaussian Mixture Model (GMM) to factorise these pre-trained embeddings for further training, thereby benefiting the \textit{cold-start} users from these pre-built social relations.}
{Our extensive experiments on three public datasets show that, \yab{in comparison to 16 competitive baselines,} our \blue{SGP model significantly outperforms the best baseline by upto 7.7\% in terms of NDCG@10.}} In addition, \iadh{we show} that \yab{SGP permits} to effectively alleviate the \textit{cold-start} problem, especially when users newly register to the system through their friends' suggestions.
\end{abstract}

\begin{highlights}
\item Social graph pre-training does improve recommendation performance

\item A Gaussian Mixture Model can effectively extract meaningful relations from the pre-trained embeddings

\item Experimental results show significant improvements using the proposed Social-aware Gaussian Pre-trained model, especially for \textit{cold-start} users

\end{highlights}

\maketitle           
\section{Introduction}\label{sec:intro}
\ioai{Deep} learning-based models have achieved \iadh{a remarkable} success \iadh{in} different \iadh{domains}\ipmm{~\citep{medi,deep_big}}. \iadh{However, although} these deep models have \iadh{a} strong expressiveness power, they \iadh{cannot} easily reach the \swext{maximal optimised solution} during the training \iadh{stage} without an effective initialisation~\citep{why_pre-train,van2020survey}. Therefore, \iadh{the} pre-training \iadh{technique} \iadh{has been \ioai{commonly} used} \xwai{to} optimise \iadh{the} deep models by \iadh{providing} them \iadh{with} an effective initialisation~\citep{why_pre-train,xin2022rethinking}. \iadh{Such a} pre-training \iadh{technique} has \iadh{been shown to lead to} state-of-the-art \iadh{performances when} \craigr{the pre-trained model is further {\em fine-tuned} to address {\em downstream}} Natural Language Processing (NLP)~\citep{gonzalez2020transformer,devlin2018bert} \craigr{or information retrieval} tasks~\citep{ma2021prop,zheng2021contextualized}. However, this effective technique \iadh{has been less studied in} recommender \iadh{systems possibly due to the limitations in the existing} datasets. \siwei{For example, in the NLP tasks, one unsupervised deep language model can be pre-trained from unlabelled texts (e.g.\ Wikipedia) and fine-tuned for a supervised downstream task~\citep{devlin2018bert,GPT-3}. \io{In contrast}, in the recommendation \swext{scenario}, each dataset contains \iadh{its} specific information about the corresponding users and items, \iadh{but} no other ground truth knowledge \iadh{such as} Wikipedia could be \iadh{leveraged} from outside the dataset \xw{to help \io{estimate} the users' preferences and items' attributes}.}

\craig{An} \iadh{existing \siweir{Neural Collaborative Filtering} (NCF) recommendation approach}~\citep{NCF} \craig{has} proposed to pre-train the recommendation model with a Multi-Layer Perceptron (MLP)~\citep{MLP}. 
Although effective, \siweir{the MLP module does not consider} \iadh{other available} auxiliary side information, \siwei{such as \iadh{the} social relations \iadh{among users}~\citep{seyedhoseinzadeh2022leveraging,elahi2021investigating} \iadh{or} \iadh{the items'} timestamps~\citep{li2020time}}, \io{therefore}
the \iadh{applied} pre-training technique of NCF \iadh{is limited in providing} the \textit{cold-start} users with \iadh{a} better initialisation.  Since the social relations among users \iadh{have been shown to be} essential \iadh{in enhancing} the recommendation performance and \iadh{alleviating} the \textit{cold-start} problem\siwei{~\citep{HGNR,camacho2018social}},
we propose to incorporate the social relations and the interaction data at the pre-training stage so that \iadh{a} better initialisation can be obtained \siweir{for} those users who have \liu{fewer} interactions.

\swai{Graph Neural Networks (GNNs), a class of deep learning \ioai{models}~\citep{GNNs_review,wu2020comprehensive}, \ioai{have been used to} aggregate \ioai{the} nodes' information from their neighbourhoods \io{so as to learn an} overall structure from \ioai{a given graph's} type of data. \ioai{Indeed}, \ioai{while} GNNs have been \ioai{previously} exploited to enhance general recommender systems\swext{~\citep{he2020lightgcn,yang2023generative,yi2023contrastive}}, they \ioai{have only been} recently studied as pre-training schemes~\citep{pretraining}. }In this work, we devise a novel Social-aware Gaussian Pre-trained model (SGP), which incorporates the \io{users'} social relations in the pre-training \iadh{stage} \siweir{and attempts to search for a \swext{relative optimised solution} based on the learned social-aware initialisation \io{during} the fine-tuning stage}. \wang{At} the first stage, we \wang{pre-train} a light GNN \wang{model} with \wang{additional} social information to
give users/items meaningful initialised embedding\swai{s}. 
\swai{Given the} {neighbourhood aggregation \iadh{property} of the GNN model,
incorporating the social relations enables socially-connected users to become closer in this latent space through the aggregation process.} 

%
\siwei{\iadh{In} the \siweir{fine-tuning} stage, we \iadh{load} the \iadh{obtained} pre-trained embeddings and re-train \iadh{the} model for further \io{recommendations}. The most \iadh{straightforward} approach \io{for} \siweir{leveraging} these pre-trained embeddings and decoding the social information is \iadh{to directly reload them. \siweir{However, }it is} essential to \iadh{note} that \iadh{the} interaction data, which will be used \iadh{in} the second stage, has \siwei{already} been exploited at the pre-training \iadh{stage}.} \siweir{Therefore,} the \ipms{direct} reuse of the interaction data might cause the overfitting problem. \ipms{To tackle the problem of data reuse, the relational knowledge distillation technique~\citep{park2019relational,gou2021knowledge} has been proposed to distil relations from a pre-trained model. \blue{The underlying intuition of the relational knowledge distillation technique is that the distillation model is encouraged to extract the essential relations from the pre-trained model, thereby avoiding a deficient performance \yab{as well as} overfitting~\citep{turc2019well}}.
\blue{Motivated by this \yab{technique of relational knowledge distillation}, we propose to distil the information from the pre-trained GNN model so that we can later reconstruct meaningful embeddings.}} 
\ipmm{Since all embeddings can be viewed as probability distributions, an intuitive solution for distilling information from those pre-trained embeddings is to follow existing works~\citep{revisit,he2020lightgcn} and use a normal distribution to model the embeddings.} \ipmm{However, those pre-trained embeddings contain prior knowledge and complex latent relations between users and items, which can hardly be modelled with a normal distribution without information loss.}
Therefore, \ipmm{during the intialisation of the fine-tuning stage}, we propose to \craigr{apply} the Gaussian Mixture Model (GMM)~\citep{GMM}, which assumes \craigr{that} all the data points are sampled from a mixture of a finite number of Gaussian distributions. By leveraging this well-developed GMM, our proposed method is devised to factorise \ipmm{the} pre-trained embeddings into a finite number of \xw{Gaussian} distributions, where this number is pre-defined and each distribution could be viewed as a specific interest of \ipmm{a group of} users or a particular characteristic of \ipmm{a set of} items.



\blue{To summarise, \yab{in this work}, we make the following contributions:}

\blue{
\noindent\textbf{$\bullet$} We devise a two-stage end-to-end social pre-trained recommendation model, SGP, which uses the GNN model to leverage social information. We show that SGP can achieve state-of-the-art performance on three real-world datasets of user-item \yab{interactions} and social relations.}

\blue{\noindent\textbf{$\bullet$} We leverage the Gaussian Mixture Model to effectively distil information from the pre-trained embeddings for the downstream recommendation task.}

\blue{\noindent\textbf{$\bullet$} Our proposed \yab{SGP} model is shown to significantly outperform \blue{16} strong baselines from the literature, while being particularly useful for \textit{cold-start} and extreme \textit{cold-start} users (newly registered users).}

The remainder of this paper is organised as follows. In Section \ref{sec:related}, we position our work in the literature. Section~\ref{sec:architecture} introduces all relevant notions used in this paper and formally defines the detailed architecture of our SGP model. The experimental setup and the results of our empirical experiments are presented in Sections \ref{sec:experiment} and \ref{sec:result}, respectively, followed by some concluding remarks in Section \ref{sec:conclusion}.

\section{Related Work}\label{sec:related}
\yab{In the following, we overview the related work from three perspectives: pre-trained models that learn general representations for various downstream tasks (Section~\ref{ss:pre-train}), graph-based recommendation models that leverage the graph structure of user-item interactions  (Section~\ref{ss:GR}), and social-aware recommendation systems, which incorporate social information into the recommendation process (Section~\ref{ss:SAR}).}

\subsection{Pre-trained \iadh{Models}}\label{ss:pre-train}
\siwei{The pre-training technique has become an emerging research topic especially in the field of NLP. Pre-trained language models such as \iadh{the} BERT~\citep{devlin2018bert} and the \iadh{more} recent GPT-3~\citep{GPT-3} models have demonstrated their robust \iadh{performance} on different \iadh{downstream NLP} tasks. \craigr{Through pre-training, a language model} can learn \swext{contextualised embeddings for tokens} from a large corpus of texts, so that these tokens can be reused for subsequent tasks with enhanced performances. \craigr{Such models can then be later \io{fine-tuned}} \iadh{for a} new downstream task, \iadh{thereby enhancing} the overall performance \iadh{of the corresponding model and outperforming other} handcrafted models. \xw{The} pre-training technique was also adopted in recommendation models.} For example, ~\citet{NCF} proposed the Neural Collaborative Filtering (NCF) model to introduce a novel deep learning-based method to \iadh{the recommender systems} community, which \iadh{has attracted} a substantial attention \iadh{from researchers} since then. The most remarkable contribution of the NCF model is that it successfully \iadh{incorporates} the multi-layer perceptron (MLP) module, which can \io{in theory} \xw{effectively} approximate \xw{various types of} \wang{prediction} functions. 
However, it is noticeable that this NCF model \iadh{also uses a} generalised matrix factorisation (GMF) module to generate pre-trained embeddings, which limits the NCF model \iadh{from incorporating} auxiliary information at the pre-training stage.  
To this end, we propose to use \iadh{instead the} GNN \iadh{technique} to replace the MLP pre-trained module due to \iadh{the former's} ability of \iadh{supporting the incorporation of} heterogeneous relations \siwei{such as \iadh{the} relations among users \iadh{as well as} the \io{users'} interaction data}. Moreover, the GNN \iadh{technique}, \iadh{which has been \siwei{initially}} devised to implement the node classification and link prediction tasks, would naturally perform better than the GMF module on aggregating similar users and items~\citep{wang2018local}. \iadh{Hence}, \iadh{compared with the GMF module}, when the GNN model is \iadh{used} for the pre-training stage, \iadh{the} embeddings of \iadh{the} socially related users can be better aggregated \iadh{in closer proximity} in the latent space. \io{Apart from} using the GMF module to pre-train on the interaction data, ~\citet{net_pretraining} introduced a linear pre-trained recommender using the network embedding method. \io{However,} their \io{proposed} model failed to leverage the multi-hops social relations (i.e. \io{a} friend's \io{friends}), which can \xw{be} seamlessly addressed by the GNN methods. The \io{study by }~\citet{pretraining} \io{is a} more recent related work \io{to ours}, \io{which} tried to tackle the cold-start problem by pre-training the recommendation model in a meta-learning setting. {However, the contribution of their work is to use the underlying structure of the user-item interaction graph, which is different from our research goal of using social information to obtain better initialised users and items' representations. Besides, existing works have leveraged other GNN pre-training and contrastive pre-training techniques for sequential~\citep{li2021pre,UPRec,xie2020contrastive} and conversational~\citep{wong2021improving} recommendations, respectively, which are distinct from our proposed general recommender SGP.}

 
\subsection{Graph-based Recommendation}\label{ss:GR}
Various graph-based recommenders\swext{~\citep{he2020lightgcn,chen2020revisiting,yu2022self,yu2022graph}} have \iadh{been shown to achieve} state-of-the-art performances \iadh{through} the development of the GNN \iadh{technique} and its variants~\citep{GCN,zhang2021mixture}.
\wang{Since the user-item interaction \iadh{data} can be \iadh{intrinsically} depicted as an interaction graph, the GNN \iadh{technique} and its variants \iadh{have} been \iadh{seamlessly} applied \iadh{in} various recommender \iadh{systems} and achieved good performances.} \iadh{For example, the} proposed \iadh{graph-based recommender} model, NGCF~\citep{NGCF}, has \iadh{been shown to outperform} many competitive baselines by incorporating \iadh{a} Graph Convolutional Network (GCN) to encode \iadh{the} collaborative signal\xw{s} and \iadh{to} model \iadh{the users'} and items' embeddings. Building on NGCF, the LightGCN \iadh{model}~\citep{he2020lightgcn} further enhanced its recommendation performance by eliminating redundant neural components from NGCF. \swext{Recently, GF-CF~\citep{shen2021powerful} \xiw{was} proposed to further enhance the performance by incorporating \xiw{a} graph filtering method. However, GF-CF is not an embedding-based model, hence it cannot be adapted to normal pre-training methods. Other variants of LightGCN, including SGL~\citep{SGL} and UltraGCN~\citep{ultragcn} have also achieved competitive performances. However, they incorporate memory-consuming data augmentation methods, which will be more challenging if side information is also considered.} Inspired by \iadh{the generalisability of LightGCN and its good trade-off between effectiveness and efficiency}, we also adopt \iadh{the} \siweir{simplified GCN~\citep{SGC}}. Moreover, we incorporate the social information into the embedding \iadh{generation} and updating process, which enables our \iadh{proposed} SGP model to encode the social relations into \iadh{the} users' embeddings. \iadh{We will show how this} auxiliary social information benefits our model \iadh{by allowing it to} obtain a better model initialisation \iadh{thereby alleviating} the \textit{cold-start} problem.

\subsection{\blue{Social-aware Recommendation}} \label{ss:SAR}
\blue{\yab{Since the early work of SoReg~\citep{SoReg}}, social relations have been \yab{shown} to be \yab {a valuable source of} side information for recommendation. Indeed, \yab{SoReg} uses social relations as a regularisation in order to enhance \yab{the} recommendation performance. \yab{As deep} learning-based \yab{recommendation} systems \yab{have developed}, researchers \yab{have also focused on how to use social relations more effectively in advanced recommendation scenarios}. \yab{This has led to the evolution of social relations usage from} traditional regularisation methods to more \yab{sophisticated} relation encoding methods~\citep{SocialGCN}. In particular, with the emergence of graph-based recommendation, social relations have become a natural choice of side information \yab{since}, within a user-item interactions graph, relations between users can be encoded as additional edges \yab{between user nodes} instead of \yab{being treated as attributes} of  entities. For example, Diffnet++~\citep{Diffnet++} \yab{incorporated} the additional social relations by adding the user-user edges into the original user-item bipartite graph. In addition, $S^2$-MHCN~\citep{yu2021self} 
\yab{has been} proposed to capture the high-order information among \yab{the} users' social relations using a hypergraph neural network augmented by the self-supervised learning technique. Similarly, SCDSR~\citep{luo2022self} is another self-supervised graph recommendation model, which \yab{built} a heterogeneous graph using the social and information domains. By doing so, SCDSR aims to leverage the high-order correlations between non-bridge users in the social domain and items in the information domain. Although \yab{overall} effective, \yab{these} existing models \yab{typically} neglect the situation when new users register \yab{to the system} through \yab{their friends' suggestions}. Compared with the existing models, our \yab{proposed} SGP model not only benefits normal \textit{cold-start} users with fewer than 5 interactions, but also provides effective recommendations to those extreme \textit{cold-start} users who have no interactions at all. }

\begin{figure}[htb]
     \centering
     \begin{subfigure}[b]{1\textwidth}
         \centering
         \includegraphics[width=0.77\textwidth]{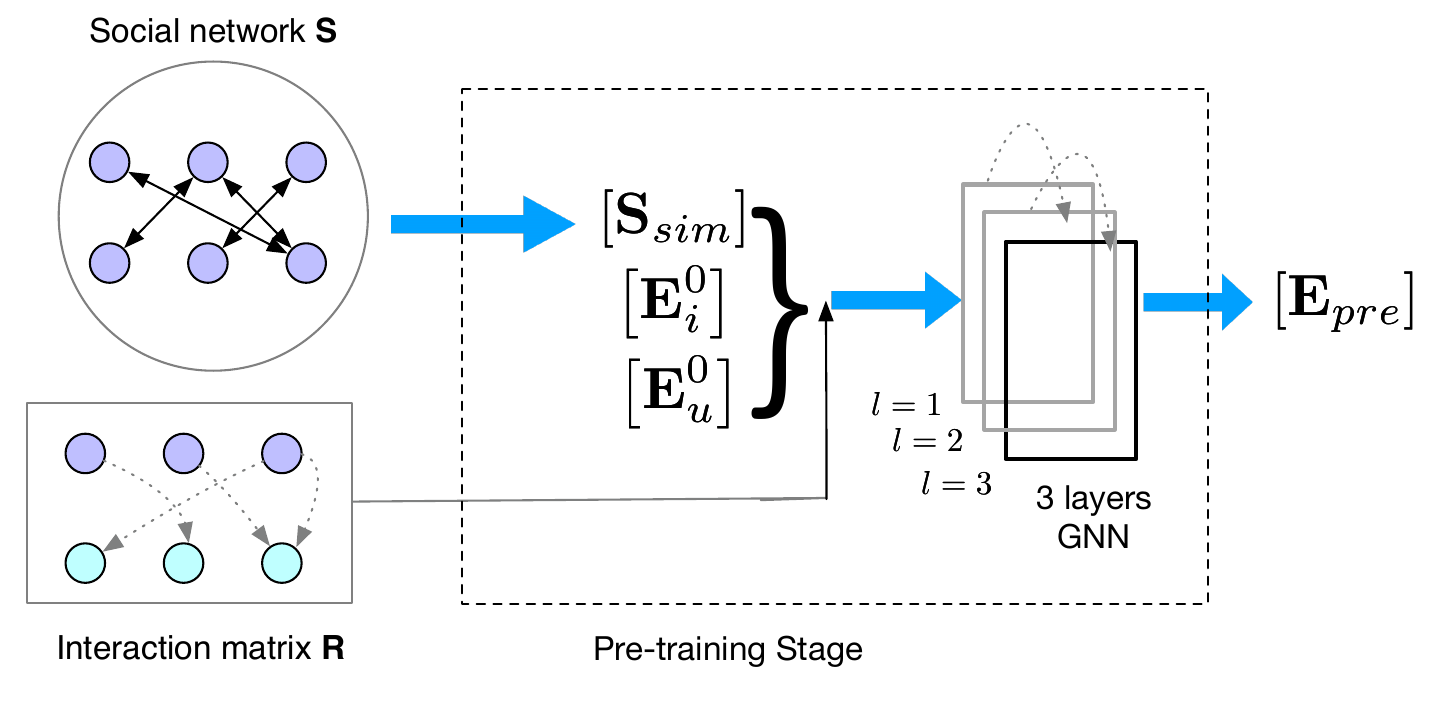}
     \end{subfigure}
     \begin{subfigure}[b]{1\textwidth}
         \centering
         \includegraphics[width=0.79\textwidth]{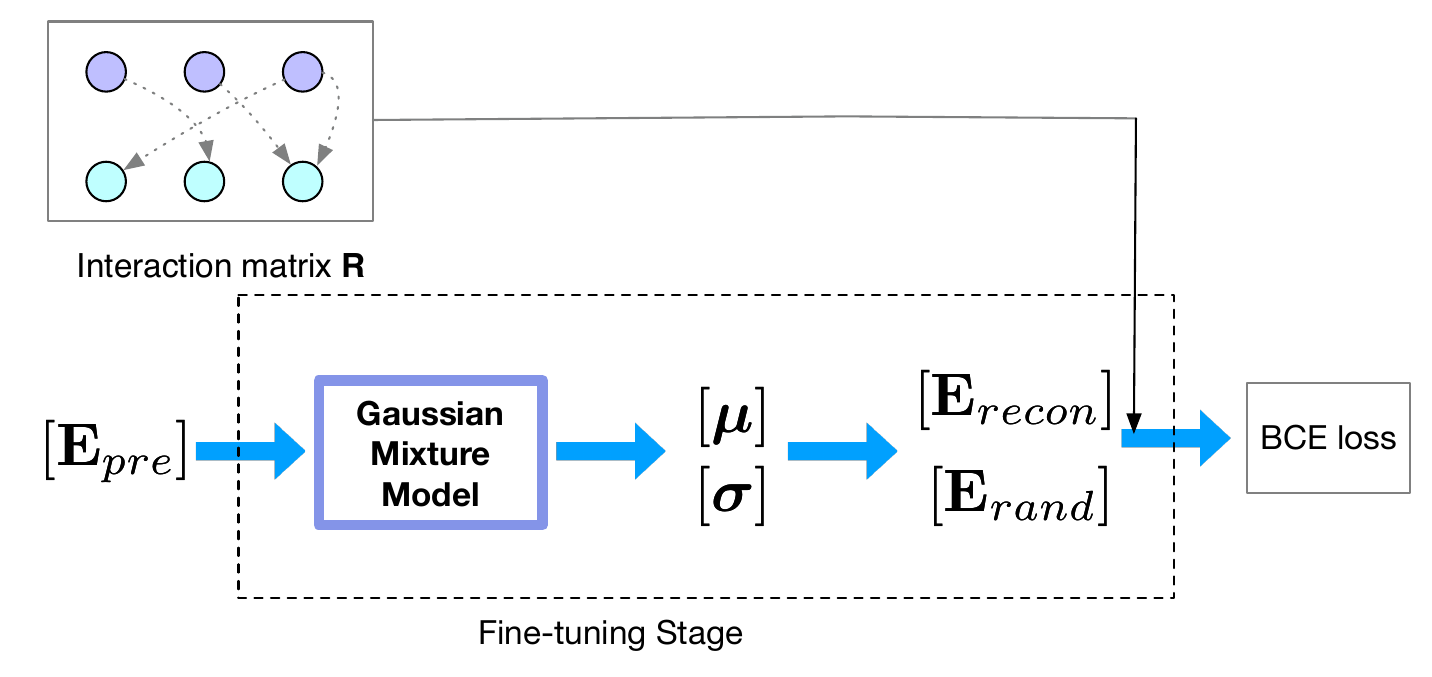}
     \end{subfigure}
     \caption{An illustration of our SGP model, where the pre-training stage and the fine-tuning stage are located above and below, respectively. \blue{In this figure, $\mymtx{E}^{0}_{u}$ and $\mymtx{E}^{0}_{i}$ are the initial embedding matrices of \yab{the} users and items, which are randomly generated. $\mymtx{S}_{sim}$ is the social similarity graph, which can be computed using \yab{Equation~(\ref{equ:sim})}. \yab{Furthermore}, $\mymtx{\mu}$ and $\mymtx{\sigma}$ contain the mean and standard deviation of the pre-trained embeddings, which are defined in Section~\ref{sec:distill}.}}
     \label{fig:overview} 
\end{figure}
\section{Model Architecture}\label{sec:architecture}
Our proposed \io{SGP} model consists of two main \iadh{stages}: 1) \iadh{a} social-aware pre-training \iadh{stage}, where a multi-layer GNN is employed to generate the pre-trained embeddings and 2) \craig{an} information distillation \iadh{stage}, where we incorporate the \swr{Gaussian Mixture Model (GMM)} to distil information from the pre-trained embeddings for the subsequent \iadh{model's} training and \iadh{generation of recommendations}.
\swr{In the following, we \ioai{first} define the tasks and \ioai{some} preliminaries in Section~\ref{ss:prelim}. \craigai{Next, Section~\ref{pre-train}}}
{\iadh{describes} how to incorporate \iadh{the} social \siwei{relations} 
and a light GNN \iadh{model} to propagate the social information into users' and items' embeddings. \craigr{Finally, in} \io{Section~\ref{ss:distill}}, we demonstrate how to employ the GMM to distil the social information from those pre-trained embeddings for the subsequent training and \iadh{the production of final} recommendations.} \swext{To clearly illustrate our model,  Figure~\ref{fig:overview} depicts the overall structure of SGP, where the upper and bottom regions describe the pre-training and fine-tuning stages, respectively.} \yab{We conclude the presentation of the SGP model with a discussion of its benefits for extreme cold-start users (Section~\ref{ss:Discussion}) as well as an analysis of its time complexity (Section~\ref{ss:TCA}).}


\begin{table}
    \centering
    \caption{Main Notations Used in this Article.}
    \begin{tabular}{P{3cm}P{8cm}}
    \hline
         Notation & Description  \\
         \hline
         $\boldsymbol{R}$ &  \multicolumn{1}{l}{the matrix of implicit feedback data}\\
         $\mathcal{U}$, $\mathcal{I}$ & \multicolumn{1}{l}{the sets of users and items}\\
         $\mymtx{S}$& \multicolumn{1}{l}{the social network matrix of all users}\\
         $\mymtx{S}_{sim}$& \multicolumn{1}{l}{the social similarity matrix of all users}\\
         $\mymtx{E}$& \multicolumn{1}{l}{the embedding matrix}\\
         $\mymtx{E}_{pre}$& \multicolumn{1}{l}{the pre-trained embedding matrix}\\
         $\mymtx{E}_{recon}$& \multicolumn{1}{l}{the reconstructed embedding matrix}\\
         $\mymtx{e}$& \multicolumn{1}{l}{the embedding vector}\\
         $\set{L}$& \multicolumn{1}{l}{the Laplacian matrix}\\
         $\mymtx{W}$ & \multicolumn{1}{l}{the learnable weight matrix}\\
         $\mu$, $\sigma$ &\multicolumn{1}{l}{the mean and standard deviation of an embedding}\\
         $\mymtx{\mu}$, $\mymtx{\sigma}$ &\multicolumn{1}{l}{the mean and standard deviation matrices of an embedding matrix}\\
         $\zeta$& \multicolumn{1}{l}{the loss}\\
         $M$, $N$& \multicolumn{1}{l}{the number of users and the number of items}\\
         $k$& \multicolumn{1}{l}{the number of Gaussian distribution}\\
         $\Theta$ & \multicolumn{1}{l}{the trainable model parameters}\\
         $\lambda$ & \multicolumn{1}{l}{the controlling factor of the $L_{2}$ regularisation }\\
         $y_{ui}$, $\hat{y}_{u i}$& \multicolumn{1}{l}{the observed interaction and the predicted interaction}\\
         $\mathcal{N}$& \multicolumn{1}{l}{the normal distribution}\\
         $\mymtx{I}$& \multicolumn{1}{l}{the identity matrix}\\
         \hline
    \end{tabular}
    \label{tab:notion}
\end{table}%

\subsection{Preliminaries}\label{ss:prelim}
In this section, we introduce the notations used across the whole article and formally define our research task. \swext{Throughout this paper, we use calligraphy typeface alphabets to denote sets (e.g. $\mathcal{U}$ is the set of users). Besides, matrices and vectors are denoted by bold letters with uppercase letters representing matrices and
lowercase letters representing vectors. In Table~\ref{tab:notion}, we summarise main notations used in this article for a fast reference.}

\siwei{\iadh{Our} task is to \iadh{highly} rank relevant items for each user given \iadh{their} historical interactions and \iadh{their} available social relations.}
We consider a recommender system with a user set $\mathcal{U}$ ($\left |\mathcal{U}  \right | = M$) and an item set $\mathcal{I}$ ($\left |\mathcal{I}  \right | = N$). Let $\mymtx{R}\in \mathbb{R}^{M\times N}$ be the user-item interaction matrix, where the content of the matrix $\mymtx{R}^{M\times N}$ corresponds to \iadh{either} explicit user ratings ~\citep{MF} or implicit feedback~\citep{BPR}. We consider implicit feedback here because it is more abundant, therefore, $\mymtx{R}_{ui}$ = 1 if the user \textit{u} has interacted with the item \textit{i}, otherwise $\mymtx{R}_{ui}$ = 0.
  \iadh{As mentioned before, social} network information is also important for improving the recommendation performance, especially when a user \iadh{does} not have enough interactions \wang{with items}. 
Let $\mymtx{S} \in \{0,1\}^{M\times M}$ be the user-user social network matrix, where the content of $\mymtx{S}$ \iadh{represents} the social connections between each pair of \iadh{users}. 


\subsection{\iadh{The} Pre-training \iadh{Stage}}\label{pre-train}

\swai{A GNN model can leverage} the \ioai{nodes'} information and \ioai{their} corresponding \ioai{relational} information in a graph by effectively aggregating information from each node's neighbours. In the recommendation scenario, each node represents either a user or an item. Therefore, suppose \iadh{that} only \iadh{the interaction} information is given, \iadh{then} each user node's neighbours \iadh{could correspond to} those items \craigr{that} have been interacted \craigr{with} by this user (\iadh{and vice-versa} for an item node). \iadh{On the other hand, when} the social network information is also \iadh{available}, then \iadh{the} user's neighbours can be his/her interacted items or friends. \wang{This friendship information is} also important for the \iadh{recommender system}, because users \swai{are more likely} to interact \iadh{with} those items \craig{that} have been \iadh{previously} interacted \iadh{with} \craig{by} their social neighbours~\citep{socialsurvey}. \craig{Hence}, \iadh{the users' available} social relations \iadh{provide useful insights for inferring their} interests and \iadh{predicting the items that they will interact with}.  

To effectively propagate \iadh{the} friends' information into each user who is socially connected, we firstly initialise each user with a randomised embedding \swr{vector} $\mymtx{e}_{u}$ to represent his/her interests. Similarly, we can assign each item with a randomised embedding \swr{vector} $\mymtx{e}_{i}$. We set $\mymtx{E}_{u}$ to be the embedding \swr{matrix} containing all latent vectors of users and $\mymtx{E}_{i}$ to \iadh{be} the embedding matrix for all items. A graph neural network can be employed here to aggregate \iadh{the} users' social information for each user node. By stacking multi-layers GNNs, we can propagate high-order connectivities of social relations from multi-hop neighbours. In our case, we \io{use} the most commonly used 3-layers GNNs to capture reasonable depths \iadh{in the} social connectivities while \iadh{avoiding} the possible over-smoothing effect of the GNN.

To \iadh{use} multi-layers GNNs\swr{~\citep{SGC}}, we rely on a well-defined Laplacian matrix $\mathcal{L}$ for the specific GNN, so that the information propagation and convolution functions can be executed effectively in a matrix multiplication form. Different \iadh{from} the NGCF~\citep{NGCF} model, which only tries to encode the \iadh{interaction} signal into both \iadh{the} users' and items' embeddings, our model focuses on the social information propagation in the pre-training \iadh{stage}. Furthermore, to \iadh{further improve} the Diffnet++ model~\citep{Diffnet++}, which only incorporates the plain social relation links, our model pre-computes the cosine similarity between each user's social relation vector~\citep{SoReg}\footnote{\io{\io{We use} the cosine similarity because it can be efficiently computed for our sparse user-user matrix. \io{In addition}, the \swr{similarities between social relation vectors} do not need estimating the magnitude, hence other similarity measures - e.g. the dot product \io{or the Euclidean} similarities, may not be appropriate}.}.
\siwei{ \iadh{These vectors constitute} one-dimensional binarised vectors indicating the social links between users.}  
Therefore, building on this advanced user-user social similarity graph $\mymtx{S}_{sim}$, the GNN function can better classify similar users. Given the user-user social graph $\boldsymbol{S} \in \{0,1\}^{M\times M}$, we can pre-compute the social similarity graph $S_{sim}$ as \iadh{follows}:
\begin{equation}
    \mymtx{S}_{sim} = \Big(\frac{\mymtx{S}\cdot \mymtx{S}^{\mathrm{T}}}{\sqrt{\diag(\mymtx{S}\cdot \mymtx{S}^{\mathrm{T}})}}\Big)^{\mathrm{T}}\cdot \frac{1}{\sqrt{\diag(\mymtx{S}\cdot \mymtx{S}^{\mathrm{T}})}},
    \label{equ:sim}
\end{equation}
where \craig{$\mathrm{T}$ represents} the transpose of a matrix and $\diag()$ \craig{computes} the diagonal matrix of the corresponding matrix. \craig{Entries of $\mymtx{S}_{sim}$ are set to 0 if the corresponding \swr{user has} no social relationships in $\mymtx{S}$}. 
Given the similarity graph $\mymtx{S}_{sim}$, we can derive its corresponding Laplacian matrix $\mymtx{\mathcal{L}}_{sim}$ as follows:
\begin{equation}
    \mymtx{\mathcal{L}}_{sim} = \diag(\mymtx{S}_{sim})^{-\frac{1}{2}}\cdot \mymtx{S}_{sim}\cdot \diag(\mymtx{S}_{sim})^{\frac{1}{2}}.
    \label{equ:laplacian}
\end{equation}
\ipmm{Next, using} the Laplacian matrix $\mymtx{\mathcal{L}}_{sim}$, we can present the embedding updating function of our proposed \iadh{SGP} model \ipmm{as follows}:
\begin{equation}
    \ipmm{\mymtx{E}_{u}^{(l)}= \mymtx{\mathcal{L}}_{sim}\cdot\mymtx{E}_{u}^{(l-1)}.}
    \label{equ:embed}
\end{equation}

Starting from a randomly initialised $\mymtx{E}^{0}_{u}$, we stack 3 layers of the GNNs given in \xw{Equation~\eqref{equ:embed}} and update the embeddings for each user correspondingly. \iadh{Following} the LightGCN model~\citep{he2020lightgcn}, we \iadh{discard} those redundant neural \swr{components} from the variant of GCN used in \swai{NGCF~\citep{NGCF}}. \swai{Indeed,} the self-connection setup, \craigr{i.e.} adding the dot product of \iadh{an} embedding with itself into Equation\xw{~\eqref{equ:embed}}, 
was initially proposed in~\citep{NGCF} to \iadh{keep} each node's original information and \iadh{to avoid being \io{possibly} overloaded with information from the nodes'} neighbours. However, this self-connection was later demonstrated in~\citep{he2020lightgcn} 
\craigai{to bring no benefit to recommendation performance; instead, it will reduce the training efficiency. }
\iadh{Hence, we choose to also remove this redundant part} in our embedding updating function following the existing work~\citep{he2020lightgcn}.

  
%
\RestyleAlgo{ruled}
\SetAlgoCaptionLayout{textbf}
\begin{algorithm}
\caption{\textbf{The pre-training stage}}
\textbf{Input:} Interaction matrix $\mymtx{R}$; Social network matrix $\mymtx{S}$\\
\noindent\textbf{Output:} Pre-trained embedding $\mymtx{E}_{pre}$.\\
Initialise embeddings  $\mymtx{E}^0$ and other learnable parameters $\Theta$;\\ 
Compute $\set{L}_{sim}$ according to Equation~\ref{equ:sim} \&~\ref{equ:laplacian};\\
\While{not early-stopped}{
$\zeta_{BCE}=0$;\\
\For{each training instance in $\mymtx{S}$}{
    Propagate social information according to Equation~\ref{equ:embed};}
\For{each training instance in $\mymtx{R}$}{
    Compute epoch loss $\nabla\zeta$ according to Equation~\ref{equ:bce};}
$\zeta_{BCE}$ $\leftarrow$ $\zeta_{BCE}$ + $\nabla\zeta;$\\
Update $\Theta$, $\mymtx{E}$;
}
\label{alg:pre-train}
\end{algorithm}

\siwei{We follow the GNN technique proposed in the aforementioned LightGCN model to update and aggregate the users' embeddings. However, different \iadh{from} LightGCN, which uses the GNN to incorporate the interaction data, we only incorporate the social information propagation. Similar with other graph-based recommendation models~\citep{GraphSAGE,NGCF,Diffnet++,ultragcn}, we \iadh{keep} the interaction data as the ground truth \iadh{for supervising} the pre-training of our model. At each training epoch, Equation~\xw{\eqref{equ:embed}} is \craig{invoked} to perform the social aggregation, after which we \iadh{use} \io{the} binary cross-entropy (BCE) loss~\citep{NCF} as the objective function:}
\begin{equation}
     \zeta_{BCE}=- \sum_{{(u,i)}
     \in \mathbf{R}} y_{u i}\cdot \log \left(\hat{y}_{u i}\right)+\left(1-y_{u i}\right) \cdot \log \left(1-\hat{y}_{u i}\right)+\lambda\left\|\Theta\right\|^{2},
     \label{equ:bce}
\end{equation}
where $y_{ui}$ is the observed \swr{interaction}, $\hat{y}_{ui}$ is the predicted \swr{interaction}, which is a dot product of the item embedding and the user embedding obtained from \craig{\xw{Equation~\eqref{equ:embed}}}, \io{while} \swr{$\Theta = \{\{\mymtx{E}_{u}^{l},\mymtx{E}_{i}^{l},\mymtx{W}^{l} \}_{l=1}^{3}\}$ denotes all trainable model parameters and $\lambda$ controls the $L_{2}$ regularisation strength to prevent overfitting.}

\swr{As a result, our pre-trained embeddings $\mymtx{E}_{pre}$
can be obtained \swai{by minimising \ioai{the objective function in} Equation~\eqref{equ:bce}.} \swext{For a better understanding, the training framework of our pre-training stage is summarised in Algorithm~\ref{alg:pre-train}.}}


\subsection{\io{The} Fine-tuning Stage} \label{ss:distill}
\siweir{\swr{After detailing the pre-training stage}, we first present the information distillation stage,
where we describe how to use the Gaussian Mixture Model to distil hierarchical relations from \io{the} pre-trained embeddings $\mymtx{E}_{pre}$. \craigr{Finally}, 
we demonstrate how to use the reconstructed embeddings $\mymtx{E}_{recon}$ for the final recommendations. \swext{Similar with the previous section, we summarise the training framework of the fine-tuning stage in Algorithm~\ref{alg:fine-tune}}. }
\subsubsection{Information Distillation \iadh{Stage}}\label{sec:distill}

By \iadh{using} \xw{Equation~\eqref{equ:bce}} \siwei{for the pre-training} \iadh{and} \xw{Equation~\eqref{equ:embed}} \siwei{for the social aggregation},
we aim to encode social information into our pre-trained embeddings \siweir{$\mymtx{E}_{pre}$. \io{The latter constitutes} the obtained embedding \swr{matrix} from optimising \xw{Equation~\eqref{equ:bce}}}. However, \iadh{it is} not obvious how \iadh{these} pre-trained embeddings can be reloaded. \iadh{Since we} have already \iadh{used} the interaction data as the ground truth \iadh{during} the pre-training stage , directly reloading these embeddings \siweir{is likely to} cause \iadh{either an} overfitting or \iadh{a marginal} improvement. Therefore, in the information distillation \iadh{stage}, we propose to distil information from \iadh{these} pre-trained embeddings. \iadh{Next}, we concatenate these distilled information at the tail of a randomly initialised embedding to add more generalisation to the final embeddings\blue{~\citep{NGCF}}. 

  Before extracting useful information from \iadh{the} pre-trained embeddings, we propose to model each user's or item's latent vector as a multi-Gaussian distribution.
\swr{This is consistent with the implementation details of existing works~\citep{BPR,revisit,NGCF,he2020lightgcn}.}
Indeed, the matrix factorisation technique can be interpreted as \iadh{the search for} the best fitted distribution for the users and items in \craigai{a} latent space. \iadh{This is} why in most cases\swr{~\citep{he2020lightgcn,revisit}}, the  embeddings  are initialised with \iadh{a} Gaussian distribution with \craig{a} \siwei{given} mean ($\mu$) and standard deviation ($\sigma $) e.g.\ $\mu = 0$ and $\sigma^2 =0.1$. \swr{As discussed in Section~\ref{sec:intro}, we \swr{expect} the pre-training stage to capture hierarchical relations between users and items\swr{~\citep{devlin2018bert}}. However, a standard Gaussian distribution \io{cannot} represent these \io{learned} complex relations from the pre-trained embeddings $\mymtx{E}_{pre}$, because its low representational power~\citep{GMM} limits its ability to convey \io{the} users' different preferences and \io{their} complex social relations, \io{thereby potentially leading to an} information loss. Hence, a mixture model is needed to leverage the possible multivariate Gaussian distributions learned from the pre-training stage and to avoid \io{any possible} information loss.}

  
%
With the aforementioned proposal, we employ a well-developed statistical analysis tool, \iadh{the} Gaussian Mixture Model (GMM)~\citep{GMM}, \craig{which} can effectively decompose a multivariate Gaussian distribution into multiple \swr{(i.e. $k$)}  Gaussian distributions, where \textit{k} is pre-defined. Specifically, GMM assumes that an observed data point $\mymtx{x}$ can be represented as a weighted sum of $k$ Gaussian densities~\citep{GMM}, calculated as follows:
\begin{equation}\label{equ:gmm_raw}
    p(\mymtx{x}|\boldsymbol{\lambda})=\sum_{i=1}^k  \mymtx{w}_i \cdot g\left(\mathbf{x} \mid \mymtx{\mu}_i, \boldsymbol{\Sigma}_i\right),
\end{equation}
where $\mymtx{x}$ is a continuous-valued feature vector, $\lambda$ represents all the learnable parameters of GMM, i.e. $\boldsymbol{\lambda} = \left\{ w_i, \mymtx{\mu}_i, \boldsymbol{\Sigma}_i\right\}$, $\mymtx{w}$ is a $k$-dimensional vector containing the weight of each Gaussian density and the sum of $\mymtx{w}_i$ is equal to 1, i.e. $\sum_{i=1}^k \mymtx{w}_i = 1$, while $\mymtx{\mu}_i$ and $\boldsymbol{\Sigma}_{i}$ denote the mean vector and the covariance matrix, respectively.


Given the feature vector $\mymtx{x}$ conditioned on the mean vector $\mymtx{\mu}_i$ and the covariance matrix $\boldsymbol{\Sigma}_i$ , we can calculate the corresponding Gaussian density as follows:
\begin{equation}
    g\left(\mathbf{x} \mid \boldsymbol{\mu}_i, \boldsymbol{\Sigma}_i\right)=\frac{\exp \left\{-\frac{1}{2}\left(\mathbf{x}-\boldsymbol{\mu}_i\right)^{\prime} \Sigma_i^{-1}\left(\mathbf{x}-\boldsymbol{\mu}_i\right)\right\}}{(2 \pi)^{D / 2}\left|\Sigma_i\right|^{1 / 2}},
\end{equation}
where $D$ is the dimension of the vector $\mymtx{x}$. In particular, the numerator is related to the Mahalanobis distance~\citep{mclachlan1999mahalanobis} (i.e. $\sqrt{\mathbf{x}-\boldsymbol{\mu}_i)^{\prime} \Sigma_i^{-1}(\mathbf{x}-\boldsymbol{\mu}_i)}$), which represents the distance between $\mymtx{x}$ and $\mymtx{\sigma}$.





Since we assume that all the users' and items' embeddings correspond to combinations of Gaussian densities, we can extract meaningful information from these embeddings by analysing each pair of ($\mymtx{\mu}_i$, $\mymtx{\Sigma}_i $), as each pair represents each user's most important preferences or each item's most important characteristics.
\begin{algorithm}
\caption{\textbf{The fine-tuning process}}
\textbf{Input:} Rating matrix $\mymtx{R}$; Pre-trained embedding $\mymtx{E}_{pre}$; Pre-defined integer $k$.\\
\textbf{Output:} The recommended list for each user.\\
Inherit $\mymtx{E}_{pre}$ for initialising embeddings;\\
Use GMM to factorise $\mymtx{E}_{pre}$ according to Equation~\eqref{equ:musigma}.\\
Randomly sample from $k$ Gaussian distributions;\\  
Generate reconstructed embeddings $\mymtx{E}_{recon}$ according to Equation~\eqref{equ:Erecon}.\\
Initialise other learnable parameters $\hat\Theta$;\\
\While{not early-stopped}{
$\zeta_{BCE}=0$;\\
\For{each training instance in $\mymtx{R}$}{
    Compute epoch loss $\nabla\zeta$ according to Equation~\eqref{equ:bce};}
$\zeta_{BCE}$ $\leftarrow$ $\zeta_{BCE}$ + $\nabla\zeta$;\\
Update $\hat\Theta$, $\mymtx{E}_{recon}$;
}
\textbf{Do recommendation to find the recommended list based on the trained embeddings according to  Equation~\eqref{equ:product}};
\label{alg:fine-tune}
\end{algorithm}
In order to extract such preferences and characteristics, we decompose the embeddings obtained from the pre-training stage in Section~\ref{pre-train} as follows:
\begin{equation}
    p(e_{u}|\boldsymbol{\lambda}) = \sum_{i=1}^{k}  \mymtx{w} _{i} \cdot g\left(\mathbf{e_{u}} \mid \mymtx{\mu}_i, \boldsymbol{\Sigma}_i\right),
    \label{equ:GMM}
\end{equation}
where $\mymtx{e}_{u}$ is the pre-trained user embedding of the user $u$ from $\mymtx{E}_{pre}$; \textit{k} is the number of Gaussian distributions as defined above in Equation~\eqref{equ:gmm_raw}, which determines how many Gaussian densities should be obtained by decomposing the pre-trained embedding. A similar equation can also be applied to an item's embedding vector.

Therefore, given the overall pre-trained embeddings $\mymtx{E}_{pre}$, we calculate all pairs of $\mu_i$ and $\Sigma_i$ as follows:
\begin{equation}
    \mymtx{\mu}, \mymtx{\Sigma} = \mathrm{GMM}({\mymtx{E}_{pre}}, k),
    \label{equ:musigma}
\end{equation}
where $\mymtx{\mu}$ and $\mymtx{\Sigma}$ are two matrices consisting of all ($\mu_i$ ,$\Sigma_i$) pairs for each user and item and the used GMM function is optimised by the EM algorithm.

After employing the GMM to those pre-trained embeddings, we \iadh{obtain} \textit{k} pairs of $\mu$ and $\Sigma$ for each user and item, \io{from} which we have enough statistical information to reconstruct the embeddings containing \iadh{the} social information encoded at the pre-training \iadh{stage}. For each user or item, we use the obtained ($\mu$ ,$\Sigma$) pairs to generate \textit{k} Gaussian distributions, where we randomly sample the same number of elements from each distribution to reconstruct socially-aware embeddings. \swr{After that, we obtain the reconstructed} embeddings $\mymtx{E}_{recon}$ \io{as follows:}
\begin{equation}
    \mymtx{E}_{recon} = \mathcal{N}(\boldsymbol{\mu}, \boldsymbol{\Sigma}),
    \label{equ:Erecon}
\end{equation}
  \swr{where $\boldsymbol{\mu}$ and $\boldsymbol{\Sigma}$ are both obtained from \xw{Equation~\eqref{equ:musigma}}.}

\subsubsection{Model Training and Recommendation}\label{sec:rec}

After obtaining the reconstructed embeddings $\mymtx{E}_{recon}$, we \iadh{use again} the BCE loss \iadh{function}~\citep{NCF} to train the model but, this time, the model is initialised with $\mymtx{E}_{recon}$ instead of a random matrix. \xw{We concatenate the reconstructed embeddings $\mymtx{E}_{recon}$ with the randomly initialised embeddings to represent \io{the} users’ preferences and items’ characteristics.}
Therefore, the model \siwei{is less likely} \iadh{to} fall into the same \swext{relative optimised solution} \siwei{within} the pre-training stage.
To recommend items of interests to a user, we compute the dot product of the \iadh{concatenated trained embeddings} of this user with \iadh{the} trained embeddings of all items in the corpus.  \swr{Hence}, \swr{our proposed \textit{Social-aware Gaussian Pre-trained} model (SGP) is devised to predict the \swr{interaction} $\hat{y}_{ui}$ between user \textit{u} and item \textit{i} as follows:}
\begin{equation}
    \hat{y}_{ui} = (\mymtx{e}_{u-recon}\parallel \mymtx{e}_{u-rand}) \odot  (\mymtx{e}_{i-recon}\parallel \mymtx{e}_{i-rand}),
    \label{equ:product}
\end{equation}
\swr{where $\parallel$ denotes the concatenation and $\odot$ is the dot product.} The obtained list of scores are \io{then} used to \io{identify the items} that \siwei{a given} user will \swr{be interested to interact with}.  

\subsection{{Discussion}} \label{ss:Discussion}
{Pre-training embeddings for the recommendation task has already been investigated in the literature. For example, to avoid the poorly performing local minima, both NCF~\citep{NCF} and CMN~\citep{cmn} apply the Generalised Matrix Factorisation (GMF) as a pre-training model to initialise the embedding of users and items. Specifically, to obtain the representations of users and items, a GMF model is trained using the weighted output of the embedding dot product:}
\begin{equation}
\hat{\mymtx{R}}_{u,v}=\swext{\mymtx{W}} \swext{\sigma}\left(\mymtx{U}_{u} \odot \mymtx{V}_{V}\right),
\end{equation}
{where $\odot$ denotes the element-wise product of vectors, $\sigma$ is an activation function and $\mymtx{W}$ is the trainable parameter.}

{However, existing models cannot leverage social relations during the pre-training stage. Hence, they may not improve the satisfaction of \textit{cold-start} users. In comparison, Our SGP model can leverage social relations using a graph-based pre-training method such that \textit{cold-start} users can benefit from friends with more interactions. 
In addition, SGP can particularly tackle the \textit{extreme cold-start} problem by inferring the preferences of those \textit{extreme cold-start} users as a combination of the preferences of their socially related friends. }

\subsection{\blue{Time Complexity Analysis}} \label{ss:TCA}
\blue{For the pre-training stage, we need to pre-compute the social similarity graph $\mymtx{S}_{sim}$ (see Equation~(\ref{equ:sim}) and the Laplacian matrix (see Equation~(\ref{equ:laplacian})), where \yab{their complexities} are $O(M^3)$ and $O(M^2)$, respectively. In addition, the complexity of \yab{the} embedding updating function (see Equation~\eqref{equ:bce}) is $O(M^2 \times |e|)$, 
 where $|e|$ is the embedding dimension.}

\blue{For the fine-tuning stage, we need to factorise $\mymtx{E}_{pre}$ using \yab{the} GMM, which has a complexity of $O(t \times k \times L \times |e|)$, where $t$ is \yab{the number of iterations for optimising the GMM function}, $k$ is the number of Gaussian distributions and $L$ is the number of training samples. Finally, the recommendation step has a complexity of $O (M \times N)$, where $M$ and $N$ are the numbers of users and items, respectively.}

 \blue{Although the pre-training stage \yab{is more time-consuming} than the fine-tuning stage, \yab{it produces reusable} pre-trained embeddings. \yab{For} a new user \yab{who joins the system} \yab{through a friend's} recommendation, the initial embedding \yab{can} be generated \yab{based on the} social relations. Since the pre-training stage does not need to be repeated as often as the fine-tuning stage, its time complexity \yab{is not a major concern} \yab{for a practical deployment}.}



\section{Datasets and Experimental Setup}\label{sec:experiment}
\iadh{To evaluate our proposed SGP model,} we perform experiments on \swai{three} public datasets: Librarything\footnote{http://cseweb.ucsd.edu/$\sim$jmcauley/datasets.html}, Epinions$^{3}$ \swai{and Yelp\footnote{https://www.yelp.com/dataset}}. \iadh{These datasets are widely used in the recommender systems community}. Librarything is a book review dataset, Epinions is a general customer \xw{review} dataset, \swai{while Yelp is a venue check-in dataset}. Table~\ref{tab:stat_all} provides the statistics of the \swai{three} used datasets. \xwai{\ioai{In the following, we} aim} to address the following research questions:

\noindent\textbf{RQ1.} Can we use the GNN model to leverage the social information and generate pre-trained embeddings for both users and items, \iadh{thereby} improving the overall recommendation performance?
    
\noindent\textbf{RQ2.} Can we employ the Gaussian Mixture Model to distil information from the pre-trained embeddings and further enhance \io{the} recommendation performance?
    
\noindent\textbf{RQ3.} Does our SGP model help in alleviating the cold-start problem, especially for those extreme cold-start users?

\noindent\textbf{RQ4.} 
\io{What is the impact of using the social relations on the pre-training stage of our SGP model?}

\noindent\textbf{RQ5.} \swext{How do the embeddings dimension and different 
ranking cut-offs affect the recommendation performances of the pre-trained recommenders?}



\begin{table}[]
\centering
\caption{Statistics of datasets.}
\begin{tabular}{c|ccc}
      & Librarything & Epinions & \swai{Yelp}\\
     \hhline{====}
     Users  & 60,243 & 114,738 & \swai{215,471}\\
     Items  & 200,422 & 34,577 & \swai{93,379}\\
     Interactions  & 930,053  & 110,671 & \swai{1,506,039}\\
     Social edges   & 110,637  & 150,859 & \swai{1,397,180}\\
     \hline
     Interaction density (\%) &0.008 &0.003 & \swai{0.007}\\     
     Social density (\%) &0.003& 0.001 & \swai{0.003}\\
\end{tabular}
\label{tab:stat_all} 
\end{table}


\io{Below, we describe the \blue{16} baselines used to evaluate the performance of SGP, the used evaluation methodology, and their corresponding experimental setup}.

\subsection{Baselines}
  
%
\swr{We compare the} performance of our SGP model \swr{to} classical strong non-neural baselines as suggested by~\citet{2019recsys}, \craigr{as well as} \io{existing} state-of-the-art neural models:
\begin{itemize}
    \item \textbf{MF}~\citep{revisit}. This is the conventional matrix factorisation model, which can be optimised by the Bayesian personalised ranking (BPR~\citep{BPR}) or the BCE losses. The regularisation includes the user bias, the item bias and the global bias. 
    \item \textbf{SBPR}~\citep{SBPR}. SBPR is a classic model, which adds the social regularisation \xw{to} the matrix factorisation method.
    \item \textbf{UserKNN} and \textbf{ItemKNN}~\citep{itemknn}. Two neighbourhood-based models using collaborative user-user or item-item similarities. 
    \item \textbf{SLIM}~\citep{ning2011slim}. This is an effective and efficient linear model with a sparse aggregation method.
    \item \textbf{NCF}~\citep{NCF}. The method is a CF model, which uses \io{a} generalised matrix factorisation method to generate pre-trained embeddings. \blue{An} MLP module is also used in NCF to capture the nonlinear features from \io{the} interactions.
    \item \textbf{NGCF}~\citep{NGCF}. NGCF is devised to employ a multi-layer GCN on top of the user-item interaction graph to propagate the collaborative signal across multi-hops user-item neighbourhoods. 
    \item \textbf{LightGCN}~\citep{he2020lightgcn}. Building on NGCF, LightGCN has fewer redundant neural components compared with the \io{original} NGCF model, which makes it more efficient and effective. 
    \item \swext{\textbf{UltraGCN}~\citep{ultragcn}. UltraGCN is a more efficient GNN-based recommender. It gains higher efficiency than LightGCN by skipping the message passing via a constraint loss.}
    \item \swext{\textbf{SGL}~\citep{SGL}. SGL leverages the self-supervised learning method to generate augmented views for nodes to enhance the model's robustness and accuracy.}
    \item \textbf{VAE-CF}~\citep{vaecf}. A state-of-the-art variational autoencoder-based collaborative filtering recommender system.
    \item \swai{\textbf{GraphRec}~\citep{graphrec}. This is the first GNN-based social recommendation \ioai{method, which} models both user-item and user-user interactions.} \item \textbf{Diffnet++}~\citep{Diffnet++}. This method is \io{a} graph-based deep learning recommender system, which uses the additional social links to enrich the user-item bipartite graph and improve the recommendation performance.   
    \item \textbf{MPSR}~\citep{liu2022multi}. This is a recent model that uses GNN to construct hierarchical user preferences and assign friends' influences with different levels of trust from different perspectives.
    \item \blue{\textbf{$S^2$-MHCN}~\citep{yu2021socially}. This is a self-supervised recommender system, \yab{which} uses a hypergraph neural network to leverage the social relations between users.}
    \item \blue{\textbf{SCDSR}~\citep{luo2022self}. SCDSR is another self-supervised graph recommendation model \yab{that} builds a heterogeneous graph using the social and interaction domains.}
    
\end{itemize}

\subsection{Evaluation Methodology}
\siwei{Following \iadh{a} common setup~\citep{BPR,deepmf}, \ioai{we use a} \xw{leave-one-out} \ioai{evaluation strategy}} to split the interactions of each dataset into training, validation and testing sets.  To speed up the evaluation, we adopt the sampled metrics~\citep{NCF,NGCF,revisit}, \io{which} randomly sample a small set of the non-interactive \craig{items} as negative items (rather than take all the non-interactive items as negatives) of the validation and testing sets, and evaluate the metric performance on this smaller set.
\blue{Here, we sample 100 negative items for each user in the testing sets for evaluation~\citep{NCF,revisit}. However, different from prior works~\citep{NCF,revisit} that only use one oracle testing set per dataset with the sampled negative items, we construct 10 different testing sets with different sampled negative items for each dataset using different random seeds, in order to reduce the evaluation bias on some specific testing negatives~\citep{krichene2020sampled}. Hence, the reported performance of each run is based on the average of the 10 testing sets}\footnote{\blue{We have also employed an evaluation methodology where any potential bias is avoided. In this evaluation, we make use of the full set of negative items. We observed similar experimental results and conclusions to those shown in Table~\ref{tab:result}. Hence, our use of 10 different testing sets enhances evaluation efficiency but also sufficiently mitigates against any potential evaluation bias stemming from the negative item sampling.}}.
\swr{In order to answer \textbf{RQ1}, we }compare our SGP model with all baselines in terms of Normalised Discounted Cumulative Gain@10 (NDCG), Recall@10 and Mean Average Precision@10 (MAP). \io{We also compare the SGP model with both its pre-training and fine-tuning stages to a variant where only the pre-training stage is used (called SGP (Pre-training)), so as to address \textbf{RQ2}}.
All models are implemented with PyTorch \iadh{using} the Beta-RecSys \iadh{open source framework}~\citep{beta_rec}. We use the Adam~\citep{Adam} optimiser \siwei{for all} the \xw{neural network} models' 
optimisations. To tune all hyper-parameters, we apply a grid search on the validation set, 
where the learning rate is tuned in~$\left \{ 10^{-2},10^{-3},10^{-4} \right \}$; \iadh{the} latent dimension in $\left \{ 32,64,128 \right \}$ 
and the $L_{2}$ normalisation in $\left \{ 10^{-2},...,10^{-5} \right \}$.  The node dropout technique is used in the NGCF, LightGCN, UltraGCN, \swai{MPSR and GraphRec} \iadh{models} \iadh{as well as} our \iadh{proposed} SGP model. \iadh{The} \xw{dropout} ratios vary amongst $\left \{ 0.3,0.4,...,0.8 \right \}$ as suggested in~\citep{GC-MC}. To control how many Gaussian distributions \iadh{are} extracted from the pre-trained embeddings, we vary the \swr{number of pre-defined multivariate Gaussian distributions} $k$ in \xw{Equation~\eqref{equ:GMM}} in $\left \{ 2,4,6,8,10 \right \}$. \siwei{\iadh{Note that} \swr{due to the limit of} the latent dimension, further \iadh{increases in} the $k$ \io{value} \siwei{might} result in less data extracted from each pre-trained embedding.}
For each $k$ value, we run our SGP model for \swext{50 times with different random seeds} and we plot the results on \iadh{the} three datasets as a box plot, \siweir{where we illustrate not only the mean values but also the variance \swr{across} different random seeds.} For a fair comparison with\swr{~\citep{NCF,he2020lightgcn,NGCF,Diffnet++}}, we set the number of neural network layers of the models including NCF, NGCF, Diffnet++, LightGCN, UltraGCN, SGL\swai{, GraphRec} and SGP to three. \swr{For \craigr{the} non-neural models, namely SBPR, MF, UserKNN, ItemKNN and SLIM, \swr{we \swr{tune} them within the same range of learning rates and $L_{2}$ normalisations used for the  neural baselines}, while for the rest of parameters we follow the same implementation details as suggested in ~\citep{2019recsys}.}

\swr{To answer \textbf{RQ3}}, we further examine the ability of our proposed \iadh{SGP} model to alleviate the \textit{cold-start} problem, especially for those \iadh{users} who \iadh{newly registered on the} sites.
In \xw{particular}, we first compare the performances of our SGP model \ioai{to} the best \iadh{performing} baseline \siweir{across different \io{groups} of users who have less than \{5, 10, 15, 20\} \io{interactions}, \io{respectively}.}
Second, to simulate the \textit{extreme cold-start} situation when a user starts using an \craigr{app} that was suggested by his/her friends, we select those users who have social relations but less than five interactions. We define these users as the \textit{extreme cold-start}\footnote{\swai{We sampled users with less than 5 interactions for the simulation because at least 3 interactions are needed for the train/valid/test set, \ioai{and in order to keep} enough users in the evaluation pool.}} users and we remove all \iadh{their interactions}, so that the \iadh{situation of newly registered users} (no historical interaction) is simulated. 
\iadh{Hence, through this defined} \textit{extreme cold-start} setup, we aim to recommend relevant items to those newly registered users \iadh{solely based on}
their social relations.

\swr{In order to tackle \textbf{RQ4,}} \siweir{we conduct an ablation study to determine the effect of the social \io{relations} in our \io{proposed} SGP model \swext{and the Diffnet++ model}. In this ablation study, we randomly drop \{20\%,40\%,60\%,80\%\} of social relations from \swext{both models} and measure the \io{resulting} recommendation performance \io{across \ioai{the} \swai{three}} used datasets, in order to determine if the performance \io{improvements are indeed} gained from the social-aware pre-training.  \swext{To answer \textbf{RQ5}, we provide a detailed analysis on the largest dataset (i.e. Yelp) \xiw{to evaluate} the performances of SGP and LightGCN on different embedding
dimensions and different cut-offs for the recommended items. }} \ioai{Additionally}, \ioai{in order to directly observe the effect of social relations in the latent space}, we use the t-distributed stochastic neighbour embedding (t-SNE) technique~\citep{tsne} to visualise \io{the} final embeddings \swai{obtained} by our SGP model, \ioai{in comparison to} \io{the} embeddings \swai{obtained} by a classic MF model.

\section{Results Analysis}\label{sec:result}
\siwei{In this section, we \iadh{report} the experimental results and \iadh{answer the \swext{five}} research questions \iadh{in turn}.}

\begin{table}[tb]
\centering
\caption{\siweir{Performances of SGP and other baselines on the three used datasets. All metrics are computed at rank cutoff 10. The best \swext{and second best performances are highlighted in boldface and underlined, respectively}; * denotes a significant difference \swr{between the performance of SGP and \io{that of} the baselines according to the paired t-test with the Holm-Bonferroni correction for $p<0.01$.} The 'Social' column indicates whether a model uses social relations or not. }}
\begin{tabular}{P{2cm}P{0.8cm}|P{0.9cm}P{0.9cm}P{0.9cm}|P{0.9cm}P{0.9cm}P{0.9cm}|P{0.9cm}P{0.9cm}P{0.9cm}}
     &Social& \multicolumn{3}{c|}{Epinions} &\multicolumn{3}{c|}{Librarything} 
     & \multicolumn{3}{c}{\swai{Yelp}}\\
     \hhline{===========}
    && NDCG  & Recall & MAP & NDCG  & Recall & MAP & NDCG & Recall & MAP\\
     NCF &\xmark& ${0.0819}^{*}$  & ${0.1662}^{*}$ & ${0.0585}^{*}$ & ${0.3132}^{*}$ &  ${0.4971}^{*}$& ${0.2304}^{*}$& ${0.2504}^{*}$& ${0.4109}^{*}$& ${0.1908}^{*}$\\
      NGCF&\xmark& ${0.0816}^{*}$  & ${0.1668}^{*}$ & ${0.0589}^{*}$ & ${0.2974}^{*}$ &  ${0.4894}^{*}$& ${0.2498}^{*}$& ${0.2378}^{*}$& ${0.3904}^{*}$& ${0.1794}^{*}$\\
     LightGCN&\xmark & ${0.0830}^{*}$  & ${0.1723}^{*}$ & \underline{${0.0615}^{*}$} & ${0.3310}^{*}$ &  ${0.5081}^{*}$& ${0.2484}^{*}$& ${0.2735}^{*}$& ${0.4304}^{*}$& ${0.2194}^{*}$\\
     UltraGCN&\xmark & ${0.0825}^*$& ${0.1700}^*$& ${0.0603}^*$ &\underline{${0.3313}^*$} &\underline{${0.5083}^*$} &${0.2480}^*$ &${0.2598}^*$ & ${0.4002}^*$&${0.2011}^*$\\
     SGL&\xmark& ${0.0831}^*$& ${0.1720}^*$ & ${0.0610}^*$&${0.3216}^*$ & ${0.4982}^*$&${0.2417}^*$ &${0.2632}^*$ &${0.4100}^*$&${0.2098}^*$\\
     VAE-CF& \xmark& ${0.0710}^{*}$& ${0.1424}^{*}$& ${0.0475}^{*}$& ${0.3003}^{*}$&  ${0.4934}^{*}$& ${0.2302}^{*}$& ${0.2100}^{*}$& ${0.3715}^{*}$& ${0.1639}^{*}$\\
     \swai{Diffnet++}& \cmark& ${0.0819}^{*}$& ${0.1678}^{*}$& ${0.0549}^{*}$& ${0.3011}^{*}$&  ${0.4873}^{*}$& ${0.2259}^{*}$& ${0.2589}^{*}$& ${0.4184}^{*}$& ${0.1988}^{*}$\\
     \swai{GraphRec} &\cmark& ${0.0810}^{*}$& ${0.1661}^{*}$& ${0.0527}^{*}$& ${0.2997}^{*}$&  ${0.4807}^{*}$& ${0.2248}^{*}$& ${0.2478}^{*}$& ${0.4097}^{*}$& ${0.1901}^{*}$\\
     MPSR &\cmark& ${0.0821}^{*}$& ${0.1692}^{*}$& ${0.0589}^{*}$& ${0.3203}^{*}$&  ${0.4927}^{*}$& ${0.2345}^{*}$& ${0.2597}^{*}$& ${0.4032}^{*}$& ${0.1925}^{*}$\\
     SBPR& \cmark& ${0.0791}^{*}$& ${0.1571}^{*}$& ${0.0509}^{*}$& ${0.2997}^{*}$&  ${0.4931}^{*}$& ${0.2300}^{*}$& ${0.2398}^{*}$& ${0.3937}^{*}$& ${0.1808}^{*}$\\
     \blue{$S^2$-MHCN} & \cmark& \blue{${0.0824}^{}$} & \blue{${0.1709}^{}$} & \blue{${0.0550}^{}$} & \blue{${0.3099}^{}$} & \blue{${0.5185}^{}$} & \blue{${0.2356}^{}$} & \blue{${0.2600}^{}$} & \blue{${0.4303}^{}$} & \blue{${0.2223}^{}$}\\
    \blue{SCDSR} & \cmark& \blue{${\underline{0.0835}}^{}$} & \blue{\underline{${0.1733}^{}$}} & \blue{${0.0604}^{}$} & \blue{${0.3276}^{}$} & \blue{${0.5020}^{}$} & \blue{\underline{${0.2501}^{}$}} & \blue{\underline{${0.2801}^{}$}} & \blue{\underline{${0.4406}^{}$}} & \blue{\underline{${0.2208}^{}$}}\\
     MF&\xmark & ${0.0720}^{*}$  & ${0.1481}^{*}$ & ${0.0484}^{*}$ & ${0.2903}^{*}$ & ${0.4893}^{*}$ & ${0.2291}^{*}$&${0.2011}^{*}$& ${0.3348}^{*}$& ${0.1698}^{*}$\\
     UserKNN &\xmark& ${0.0752}^{*}$ & ${0.1678}^{*}$& ${0.0497}^{*}$& ${0.3123}^{*}$&  ${0.4987}^{*}$& ${0.2345}^{*}$&${0.2297}^{*}$& ${0.3797}^{*}$& ${0.1758}^{*}$\\
     ItemKNN &\xmark& ${0.0743}^{*}$& ${0.1667}^{*}$& ${0.0486}^{*}$& ${0.2977}^{*}$&  ${0.4872}^{*}$& ${0.2139}^{*}$&${0.2238}^{*}$& ${0.3709}^{*}$& ${0.1712}^{*}$\\
     SLIM &\xmark& ${0.0719}^{*}$& ${0.1522}^{*}$& ${0.0497}^{*}$& ${0.2918}^{*}$&  ${0.4821}^{*}$& ${0.2298}^{*}$&${0.2098}^{*}$& ${0.3407}^{*}$& ${0.1766}^{*}$\\
     \siwei{SGP (Pre-training)} &\cmark& 0.0725  & 0.1498 &0.0497 & 0.2953 &0.4913 &0.2284& 0.2201& 0.3897& 0.1798\\
     \hline
     SGP &\cmark & \textbf{0.0876}  & \textbf{0.1794}& \textbf{0.0657} &\textbf{0.3569} &  \textbf{0.5431} & \textbf{0.2647}& \textbf{0.2972} & \textbf{0.4631}& \textbf{0.2347}\\
     \%Improv.& &4.9 &3.5&6.8&7.7 & 6.9 & 5.8& 6.1& 5.1& 6.3\\
\end{tabular}
\label{tab:result} 
\end{table}

\begin{table}[tb]
\centering
\caption{NDCG@10 \ioai{performances} of our SGP model \iadh{in comparison to} the LightGCN baseline across different user groups where n is the number of users' interactions; \siwei{* denotes a significant difference \xwai{versus} \siwei{LightGCN} (paired t-test, p$<$0.01).} 
}
\begin{tabular}{c|c|ccccc}
&&&NDCG@10&&&\\
Dataset& Model & n=5& n=10& n=15& n=20 & n=all \\ 
\hhline{=======}
\multirow{3}{4em}{Epinions}     & SGP & $0.0763^{*}$ &  $0.0812^{*}$ & $0.0857^{*}$& $0.0864^{*}$& $0.0876^{*}$ \\ 
                              & LightGCN & 0.0708 & 0.0751 &  0.0810 & 0.0818& 0.0830\\  
                              &\%Improv. & 7.77& 8.12& 5.80& 5.62& 5.50\\
\hline

\multirow{3}{5.5em}{Librarything} & SGP & $0.3014^{*}$ & $0.3354^{*}$ & $0.3410^{*}$ & $0.3554^{*}$ & $0.3569^{*}$\\ 
                              & LightGCN & 0.2631  & 0.2821 & 0.3101  & 0.3275& 0.3310\\ 
                              &\%Improv. & 14.5& 15.9 & 9.96& 8.52& 7.80\\
\hline

\multirow{3}{4em}{\swai{Yelp}}     & \swai{SGP} & $0.1987^{*}$ &  $0.2435^{*}$ & $0.2669^{*}$& $0.2848^{*}$& $0.2972^{*}$ \\ 
                              & \swai{LightGCN} & 0.1671 & 0.2086 &  0.2381 & 0.2590& 0.2735\\  
                              &\swai{\%Improv.} & 18.9 & 16.7&12.1 & 9.97& 8.71\\
                              
\end{tabular}
\label{tab:cold} 
\end{table}

\begin{table}[tb]
\centering
\caption{\iadh{Performances of} SGP \siwei{(Pre-training)} on the \textit{extreme cold-start} users in comparison \iadh{with the random and popularity-based baselines}. \craig{In the table, SGP is the only approach where interactions are used.}
\siwei{* and $\uparrow$ denote significant differences compared to the \craig{random and popularity} \craig{baselines}, respectively (paired t-test, p$<$0.01).}} 
\begin{tabular}{P{1.6cm}|P{1cm}P{1cm}P{1cm}|P{1cm}P{1cm}P{1cm}|P{1cm}P{1cm}P{1cm}}
     & \multicolumn{3}{c|}{Epinions} &\multicolumn{3}{c}{Librarything} & \multicolumn{3}{c}{\swai{Yelp}}\\
    & NDCG  & Recall & MAP & NDCG & Recall & MAP & NDCG & Recall & MAP\\
     \hhline{==========}
     Random & 0.0676  & 0.1334 &0.0403 & 0.0998 &  0.1938& 0.0707 & 0.0887& 0.1806& 0.0615\\
      Popularity & 0.0712 & 0.1448 & 0.0423 & 0.1693 & 0.3011 & 0.1082 & 0.1937& 0.3219&0.1287\\
     SGP & $0.0870^{*\uparrow}$  & $0.1691^{*\uparrow}$ & $0.0589^{*\uparrow}$ & $0.3324^{*\uparrow}$  & $0.5134^{*\uparrow}$ & $0.2958^{*\uparrow}$ &
     $0.2972^{*\uparrow}$  & $0.4631^{*\uparrow}$ & $0.2347^{*\uparrow}$\\
     \hline
     SGP (Pre-training) & $0.0728^{*\uparrow}$ & $0.1583^{*\uparrow}$ &$0.0450^{*\uparrow}$ & $0.2029^{*\uparrow}$ & $0.3360^{*\uparrow}$ & $0.1382^{*\uparrow}$ & $0.2279^{*\uparrow}$ & $0.3469^{*\uparrow}$ & $0.1488^{*\uparrow}$\\
\end{tabular}
\label{tab:non_exist} 
\end{table}

\subsection{RQ1: Pre-trained Recommendation Performances}
  
%
\swr{In order to answer \textbf{RQ1}, we use Table~\ref{tab:result} to report }the overall performance of our SGP model in comparison to \swext{13}
\io{other} baselines and the pre-training stage \iadh{(the first stage only of the SGP model})
in terms of 3 \siwei{different} metrics, \siwei{namely NDCG, Recall and MAP}.
\swr{Comparing the performance of SGP (Pre-training) with other baselines, we can conclude that the pre-training stage itself cannot outperform all baselines.}
However, through the information distillation stage when we use randomly initialised embeddings \swr{concatenated} with Multivariate Gaussian distributions extracted from the pre-trained embeddings, our SGP model achieves the best performance, constantly and significantly \iadh{outperforming} all other baselines in terms of all metrics on \swai{three} \iadh{used} datasets. \iadh{These results} demonstrate that solely employing the GNN \iadh{model} with the available social relations is not \iadh{sufficient} to enhance the \iadh{recommendation} performance. This is likely because the social information should not be \io{considered equally to} the interaction information, \iadh{since the interaction information} \iadh{makes the actual} ground truth when inferring \iadh{the} users' main preferences and next items of interest.
By reusing these pre-trained embeddings \swr{concatenated with} randomly initialised embeddings, our SGP model can \io{markedly} \swr{and significantly} enhance the \iadh{recommendation} performance. \swr{It is of note that the performances of all the evaluated models on the Epinions dataset are lower than on \io{the} Librarything \swai{and Yelp datasets}. However, these performances are in line with those reported in the literature (e.g. NDCG@10 $\approx$ 0.3 on the Librarything dataset~\citep{LT1,LT2,LT3}; NDCG@10 $<$ 0.1 on the Epinions dataset~\citep{EP2}). These differences may be explained by the differing densities of user-item interactions in the \io{used} datasets (see Table~\ref{tab:stat_all}).} 
Additionally, by comparing other graph-based models, we observe that those more recent models such as SGL and UltraGCN do sometimes outperform the LightGCN model. However, LightGCN can still outperform SGL and UltraGCN
for most of the times as shown in Table~\ref{tab:result}.
This observation is \blue{likely} related to our used data split, where we use 10 different testing sets to avoid the oracle testing set. In other words, our results \blue{suggest} that \blue{the recent graph-based models have not achieved consistent and robust} \blue{improvements} over the LightGCN model. \blue{In addition to those graph-based baseline models, our SGP also significantly outperforms other social-aware recommender systems, including $S^2$-MHCN and SCDSR. This indicates that a social graph pre-training technique is more effective than the self-supervised learning technique on the recommendation task.} 
\swr{Overall, in answer to \textbf{RQ1}, we can conclude that using the GNN model to leverage the social relations and generate pre-trained embeddings can improve the recommendation performance compared with SGP (Pre-training) and \blue{16} competitive (neural and non-neural) baselines.}





\subsection{RQ2: GMM Information Distillation}
    
\begin{figure}[tb]
    \centering
    \includegraphics[width=16cm]{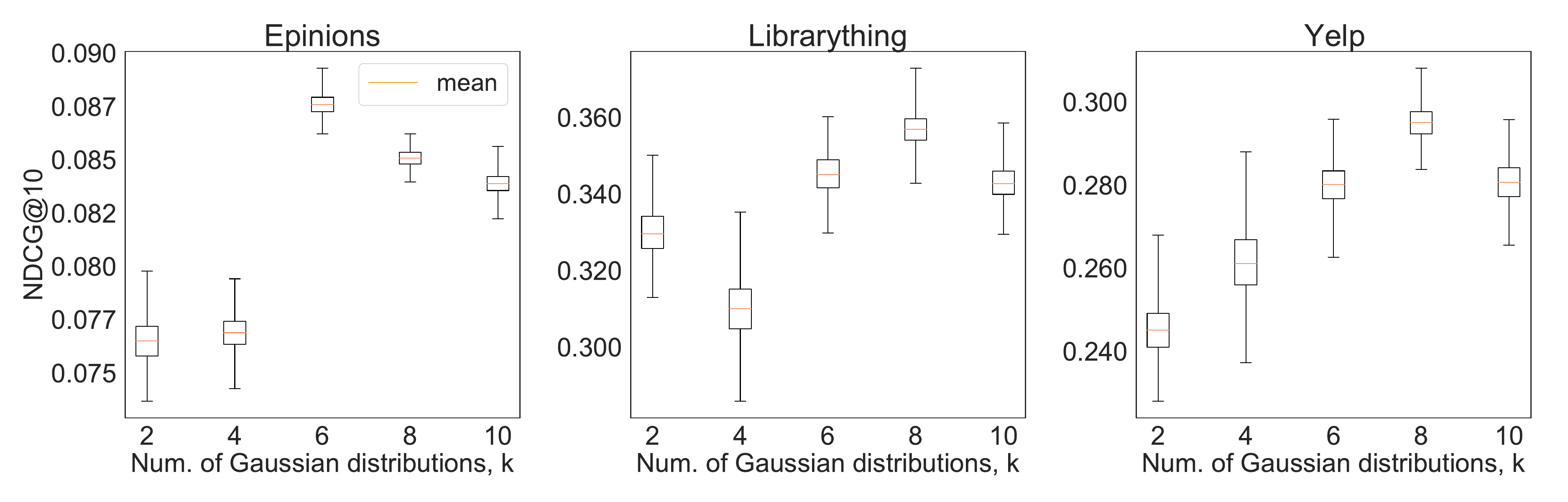}
    \caption{Max/Min/Mean values of NDCG@10 for \xwai{SGP} on 3 datasets with different number of Gaussian distributions.}
    \label{fig:num_gau} 
\end{figure}

\swr{To \craigai{address} \textbf{RQ2}}, we \iadh{show a} box plot of our SGP model on \iadh{the 3 used} datasets across different number of pre-defined multivariate Gaussian distributions, $k$, in terms of NDCG@10, 
where for each $k$ value, the model is trained and evaluated \swext{50 times with different random seeds}. In Figure~\ref{fig:num_gau}, the max and min values for each set of \iadh{experiments} are shown as two bars at the top and bottom of each box, respectively. The mean value of each set of \iadh{experiments} is shown as \iadh{an} orange line lying in the middle of each box. \iadh{We also} report the best mean for each dataset in Table~\ref{tab:result} (i.e. $k=6$ for the Epinions dataset and $k=8$ for the Librarything \swai{and Yelp} datasets ). Figure~\ref{fig:num_gau} \iadh{shows} that our SGP model only achieves better \iadh{performances} when $k$ is larger than 4, \iadh{whereas} \swr{for \swai{all} datasets} when $k=2$ or 4, the SGP model \iadh{has a} lower performance than several baselines. This can be explained \iadh{by the fact that the} users' preferences are \iadh{hard to be estimated with simple distributions}. \craigr{Indeed,} \siwei{usually \iadh{the users' preferences are \iadh{formed by} combinations of distributions}, which \iadh{cannot} be \iadh{easily} factorised with 2 to 4 factors.} 
Therefore, \iadh{a} small number of Gaussian distributions \iadh{is} not \iadh{sufficient} enough to represent \iadh{the} users' preferences. However, we also observe a performance degradation when $k$ is too large. This is likely because the latent dimension \siwei{has a limited size (usually up to a few hundreds)},
while each reconstructed embedding is a sample from the multiple extracted Gaussian distributions. Therefore, when $k$ \iadh{becomes} larger, elements sampled from each distribution become fewer, \io{thereby leading to a loss in the accuracy of the representation of its intended original factor}.
For example, when the latent dimension is 100, if $k=10$ is applied, only 10 elements are sampled from each Gaussian distribution.
Moreover, when k is larger, the performance of our SGP model is relatively stable. This demonstrates that our SGP model \iadh{is} effective \iadh{in distilling} information from the pre-trained embeddings given that enough Gaussian distributions are employed \swr{i.e. when $k$ is sufficiently large, the model stabilises and shows less variance.} In addition, we observe that the performance of SGP in terms of NDCG@10 varies across different datasets. For example, when $k=4$, SGP is relatively less effective than the other configurations on the Librarything dataset while SGP is dramatically improved when $k\geq 6$ on the Epinions dataset. As shown in Table~\ref{tab:stat_all}, different datasets have different social densities leading to different structures of social networks. Therefore, these structures of the social networks can affect the usefulness of the social relations thereby also affecting the performance of social-aware recommenders such as SGP. \swr{Overall, in answer to \textbf{RQ2}, \io{we can conclude} that the GMM can be \io{used to effectively} distill information from the pre-trained embeddings. \io{We also} suggest preferable $k$ values, which can be used to enhance the recommendation performance. }


\subsection{RQ3: \textit{Cold-start} Performances}\label{sec:cold}
\swr{To address \textbf{RQ3}, in Table~\ref{tab:cold}, we }examine \io{the} \iadh{performance} \io{of our SGP model} for \siweir{different \io{groups} of users  who have less than \{5, 10, 15, 20\} \io{interactions}, \io{respectively}}, in comparison to the best baseline, LightGCN, in terms of NDCG@10. \iadh{From the table, we note} that our SGP model \iadh{overall \liu{significantly} outperforms LighGCN}, while users with less than 10 interactions \iadh{particularly} benefit from our model compared with the \siweir{other \io{groups} of users}. 
\iadh{Overall, it is} reasonable to observe that \textit{cold-start} users benefit more from our model because when their interaction information is too sparse, incorporating more social information will \io{likely} enable the \iadh{SGP} model to predict their possible unknown preferences. \siweir{However, users with \craigr{sufficient interactions} tend to have their preferences \io{accurately} captured by \io{the} recommender systems,} therefore adding more social relations may not be beneficial for them. Indeed, from Table~\ref{tab:cold}, we \io{observe} that there is a \siweir{clear \io{decrease} in the \io{reported} percentage improvement \io{when we consider} the group of users who have less than 10 interactions \io{in comparison to those users who have} more than 15 interactions.}


%
Table~\ref{tab:non_exist} \iadh{shows a} comparison \iadh{of} our SGP model \iadh{with} \liu{\iadh{a} random recommender and \iadh{a popularity-based} recommender \siweir{for the \textit{extreme cold-start} users} case. \iadh{The \io{random and popularity-based recommenders} are} two commonly used baselines when no interaction data is available.}
\iadh{Here,  we aim to \io{simulate}} the situation when users register \iadh{to} an App or \iadh{a Web service following} the suggestions of \iadh{their} friends. In this case, the model only knows \iadh{about} \iadh{the} users' friends while \iadh{it} does not have access to the historical interactions. Instead of making random recommendations \liu{or \iadh{only} recommending popular items}, our SGP model generates embeddings by constructing multivariate Gaussian distributions \iadh{by} evenly sampling elements from their \iadh{friends'} embeddings, which is also \swr{produced} \iadh{by} the pre-training stage \siwei{of our SGP model}. By comparing our proposed SGP \iadh{model with its first pre-training stage only} (denoted by \siwei{SGP (Pre-training))} with \iadh{both} \liu{baselines} and the \iadh{full} SGP model \iadh{for} the \textit{extreme cold-start users}, 
we find that \siwei{SGP (Pre-training)} 
\siwei{significantly \iadh{outperforms}} 
\liu{\iadh{both} the random and the \iadh{popularity-based} recommenders}. \iadh{On the other hand, it} \iadh{is} reasonable and \iadh{natural} that when no interactions \iadh{are observed} and no training is \iadh{conducted}, \siwei{SGP (Pre-training)} is far worse than the \siwei{full} SGP  \iadh{model}. \wang{However, SGP (Pre-training) significantly outperforms both the random and the popularity-based recommenders on the Librarything \swai{and Yelp} datasets and is comparable to the results of SGP on the Epinions dataset.}
\iadh{Overall}, in answer to \textbf{RQ3}, we can conclude \swr{that our proposed method SGP is effective at tackling the \textit{cold-start} problem and is }
\iadh{particularly} useful \io{in alleviating} the practical \swr{extreme cold-start} issue.


\subsection{RQ4: Impact of Social Relations}\label{sec:ablation}
\begin{figure}[htb]
     \centering
     \begin{subfigure}[b]{1\textwidth}
         \centering
         \includegraphics[width=0.30\textwidth]{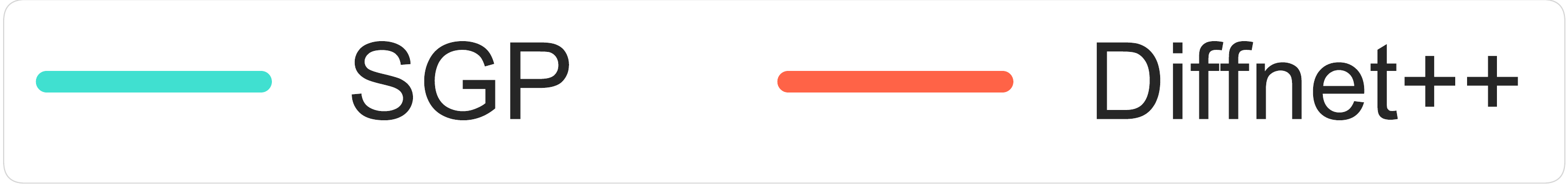}
     \end{subfigure}
     \begin{subfigure}[b]{1\textwidth}
         \centering
         \includegraphics[width=1\textwidth]{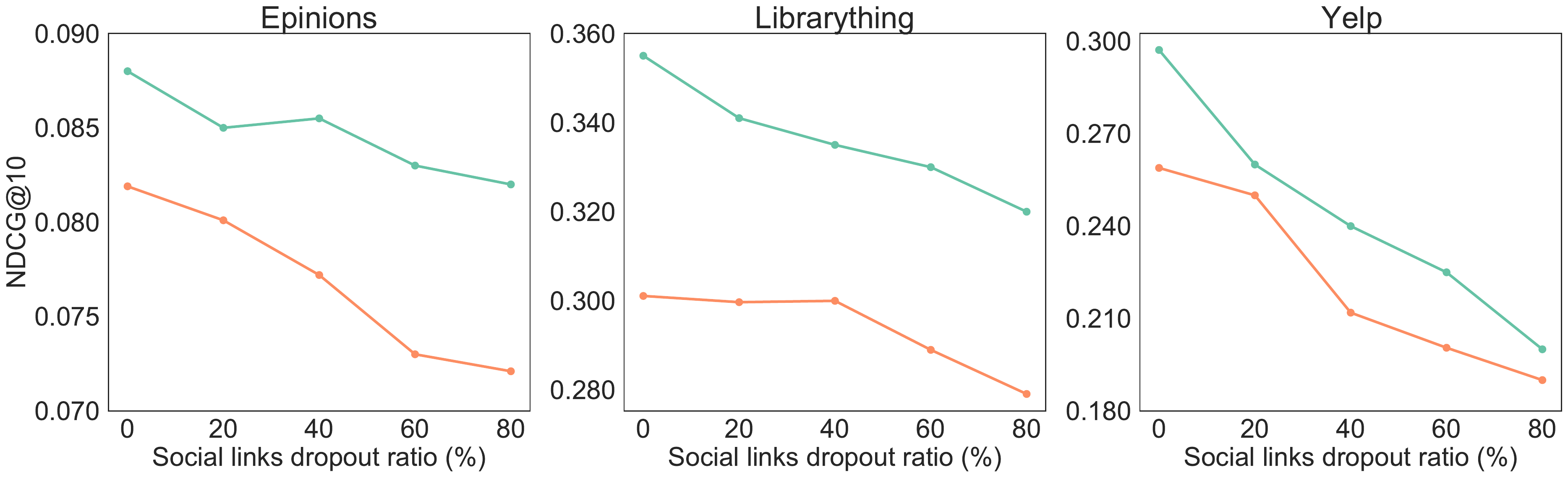}
     \end{subfigure}
     \caption{\siweir{An ablation study of \swext{performances of SGP and Diffnet++} (different proportions of social relations are masked out).}}
     \label{fig:ablation}
\end{figure}

\swr{In order to determine the effect of social relations and to answer \textbf{RQ4}, we conduct an ablation study \io{where} we randomly dropout different proportions of social relations.} \siweir{Figure~\ref{fig:ablation} \io{shows} how the performance of our SGP model \swext{and Diffnet++ model are} affected when different proportions of social relations are randomly masked out. From this figure, we can clearly observe \craigr{a trend that when} more social relations are masked during the pre-training, the \io{more the} recommendation performances of \swext{SGP and Diffnet++ are} \craigr{ degraded} across \swai{three} datasets. This trend reveals that  \io{the social relations do indeed} help the SGP model to \io{achieve} a better pre-training \io{thereby} enhancing the final recommendation performance. However, \io{we also observe} some variance in the performance on the Epinions dataset, compared with the consistent decline of performance on the Librarything \swai{and Yelp} datasets. This is because the raw data of the Epinions \io{dataset} provides bidirectional social relations i.e.\ both \io{the} `trust' and `trustedby' relations are given. Since our \io{current} SGP model \io{cannot} distinguish \io{between} these bidirectional relations, for the sake of simplicity, we unify these two types of relations as one unidirectional social network to fit our implementation. Although \swr{unifying the bidirectional relations does bring an} overall performance improvement to SGP over other baselines, this unifying method itself is \io{not optimal} and can possibly induce noise, because \swr{the social influences are not bidirectionally equal. \swext{By comparing the Diffnet++ model with SGP across three datasets over different dropout ratios, we observe consistent performances improvement from our proposed model, which further justifies our previous results.}
Therefore, from \io{our conducted} ablation study, in answer to \textbf{RQ4}, we can conclude that \swr{using the social relations on the pre-training stage can help enhance the recommendation performance of our SGP model. }
Furthermore, we \io{postulate} that the performance can be further \io{enhanced} by enabling our current SGP model to distinguish \io{among} bidirectional social relations. \io{We leave the adequate integration of bidirectional social relations into our SGP model to future work}.}}


\begin{figure}[htb]
     \centering
     \begin{subfigure}[b]{1\textwidth}
         \centering
         \includegraphics[width=0.28\textwidth]{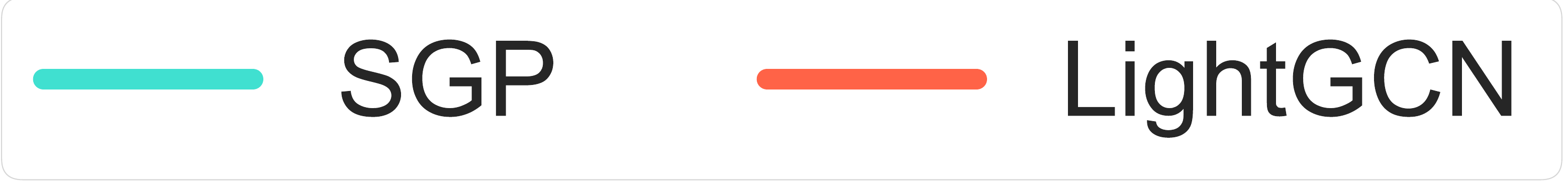}
     \end{subfigure}
     \begin{subfigure}[b]{0.41\textwidth}
         \centering
         \includegraphics[width=1\textwidth]{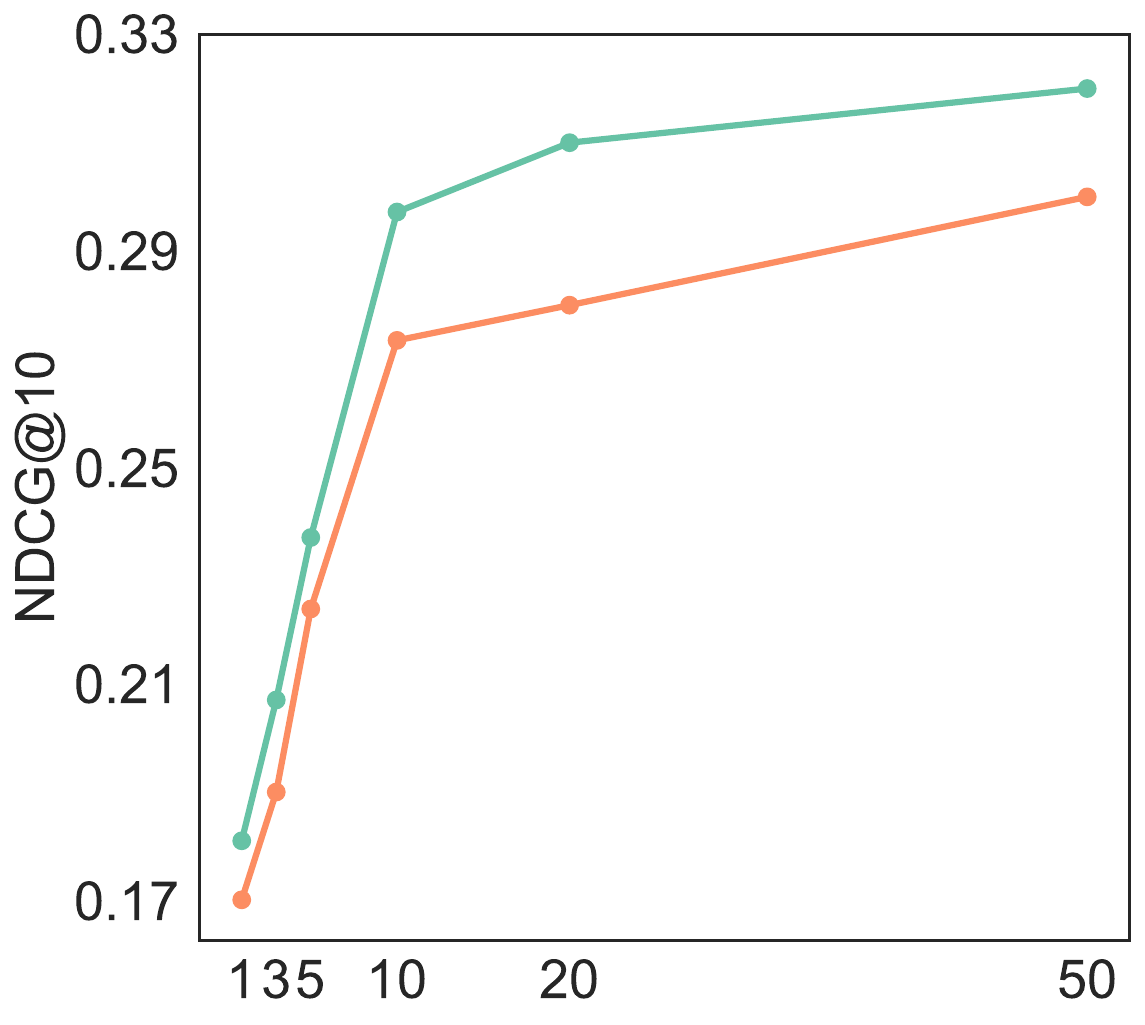}
         \caption{\footnotesize Cut-offs}
         \label{fig:cut}
     \end{subfigure}
     \begin{subfigure}[b]{0.4\textwidth}
         \centering
         \includegraphics[width=1\textwidth]{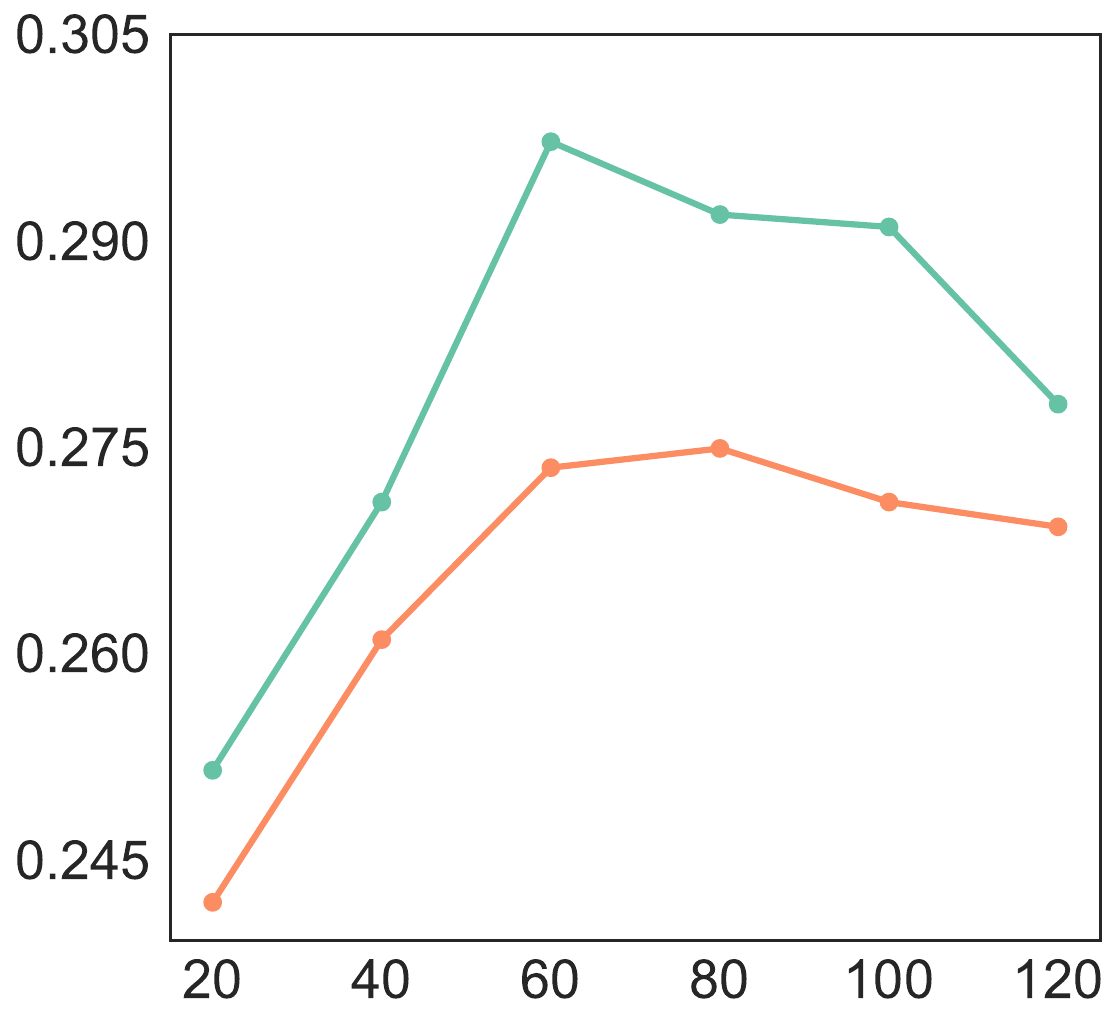}
         \caption{\footnotesize Dimensions}
        \label{fig:dim}
     \end{subfigure}
     \caption{\swext{A performance comparison between SGP and LightGCN over different dimensions and different cut-offs on the Yelp dataset.}}
\end{figure}

\subsection{RQ5: Hyperparameter Analysis}\label{sec:hyper}
\swext{In this section, we aim to answer \textbf{RQ5} by examining how the performance of SGP and that of the second-best baseline LightGCN are affected by different recommendation cut-offs and embedding dimensions. First, we plot when the recommendation lists are generated with different cut-offs in Figure~\ref{fig:cut}. 
From this figure, we can observe that our SGP consistently outperforms LightGCN across different cut-offs. Specifically, SGP mainly surpasses LightGCN for larger cut-offs (i.e. when cut-offs $\geq 10$). This is due to the fact that we consider social relations as side information, and they are only leveraged during the pre-training stage. As a result, those items that are easy to predict will be preserved as top-ranked items, while social relations play an important role for our SGP model in obtaining a higher effectiveness at lower rank cutoffs.
For example, in the venue recommendation scenario, users may visit venues suggested by their friends when travelling to different countries. When these venues are in the test set, a general recommender relying only on the interaction information will unlikely rank these venues at the top of the ranking list for such users. Our proposed SGP model is likely to benefit this case if the users’ friends have visited/liked these venues before. Specifically, SGP gives higher scores to those venues visited/recommended by each user’s friends. As a result, those venues lowly ranked by other recommenders will have a higher chance to appear in the top-10/top-20/top50 ranking lists as shown in Figure~\ref{fig:cut}.}
Here, we must emphasise the difference between the results shown in Figure~\ref{fig:cut} and Table~\ref{tab:cold} to avoid a possible confusion. In Table~\ref{tab:cold}, we have reported performances across different user groups defined by the number of historical interactions of each user.
This is different from what we plot in this section, which is based on different cut-offs. 
Figure~\ref{fig:dim} shows how the NDCG@10 measure is affected when different embedding dimensions are applied to SGP and LightGCN. This figure demonstrates that our SGP can bring consistent improvements over the baseline for different embedding dimensions. To conclude on \textbf{RQ5}, our SGP model can constantly outperform the strong LightGCN baseline when different hyperparameters are applied.

\subsection{\swext{The Embedding Visualisation}}

\begin{figure}[htb]
    \centering
    \includegraphics[width=16cm]{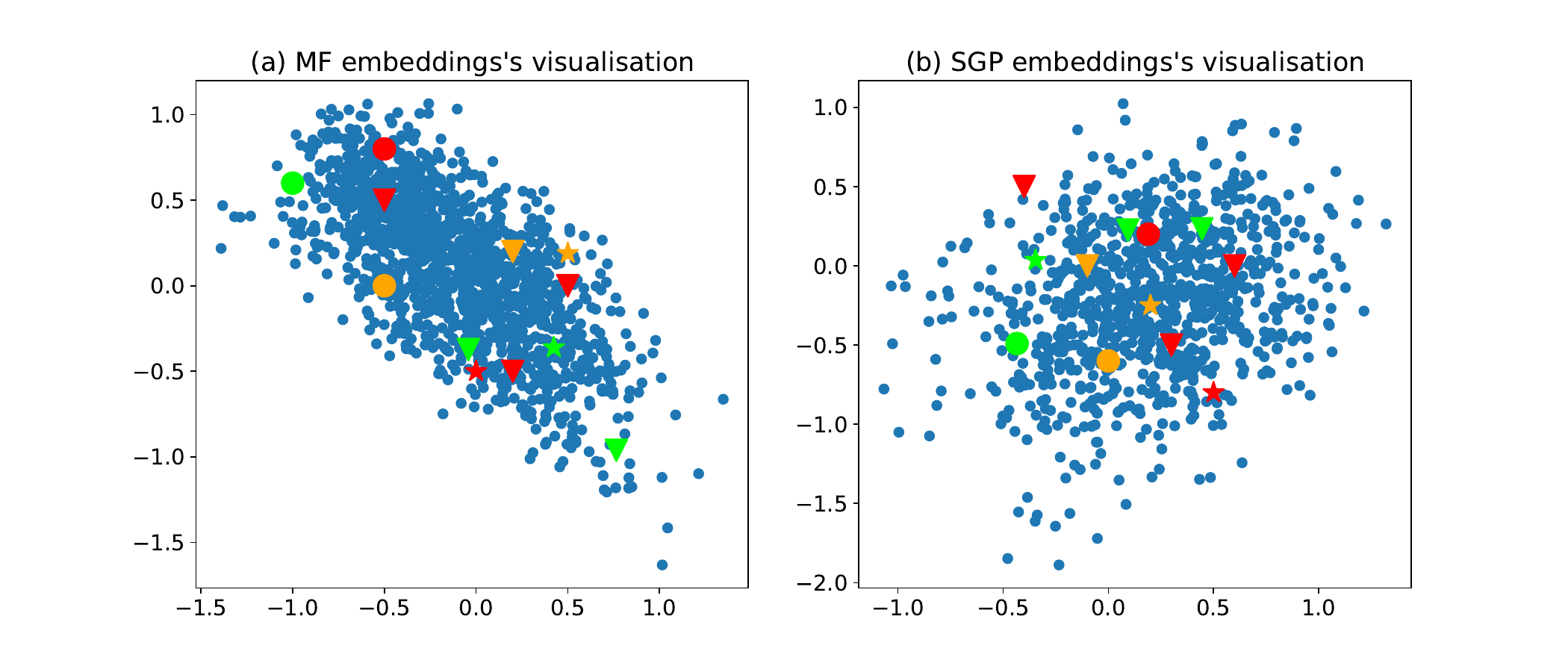}
    \caption{\siweir{The t-SNE plot of all users' embeddings of the Librarything dataset obtained from the MF model (a) and our SGP model (b), where \swr{the red dot represents an anchor user, the red triangles are this user's friends and the red start is the target item. The similar configuration is also applied to another two users with their corresponding friends and target items, which are plotted with the colour of green and orange, respectively.}}}
    \label{fig:tsne}
\end{figure}

\swr{In this section, we aim to analyse how our SGP model affects the \io{users'} embeddings in the latent space, compared to the embeddings obtained from a \io{classic} MF model that does not encapsulate social relations. }\siweir{We visualise all \io{users'} embeddings \swr{of the Librarything dataset\footnote{\swr{The Librarything dataset \io{is} a less sparse dataset with a higher or equal social density compared to the Epinions \swai{and Yelp} datasets \io{therefore, for illustration purposes}, we have more users to choose from. \io{However, note that we do nevertheless} observe similar \io{trends} on the Epinions \swai{and Yelp} datasets}}}, trained by the SGP model in comparison with embeddings trained by the MF model\footnote{\swr{The MF model is chosen because \craigr{it is} also \craigr{an} embedding-based method and \craigr{is not socially aware}, and \craig{therefore} can offer us a clear comparison between a social-aware model and a non-social model.}} using the t-distributed stochastic neighbour embedding (t-SNE) approach. \craigr{Figure~\ref{fig:tsne}(a) shows the t-SNE for MF, while Figure~\ref{fig:tsne}(b) \io{provides} the t-SNE for SGP. In both plots}, we \swr{highlight} 
three anchor users (\craigr{represented as} yellow/green/red dots), \craigr{along with} their corresponding friends (triangles) and \swr{their} target items (\craigr{stars}). \swr{Both the green and orange anchor users are fortunate to have their friends close \craig{to} their target items, hence, these two anchor users are pulled closer to their target items, as shown in  Figure~\ref{fig:tsne}(b). \craig{In contrast}, in Figure~\ref{fig:tsne}~(a), these two anchor users are clustered far apart from their target items by MF, due to the \io{fact that} social relations \io{are not} considered by MF. For the red user's case, he/she has a dissimilar friend, \craigr{who} is located relatively far away from the target item and \swr{his/her} friends. Our SGP model can still handle this case by relocating the red user \swr{to} the space between this dissimilar friend and two \swr{other} similar friends, \io{thereby bringing} this user \io{closer} to the target item. \io{Through the} provided three examples of users, \io{we illustrated} different situations \io{where} users \io{might} possibly benefit from our SGP model, \io{thereby improving the recommendation performances as observed in the reported results across three datasets}. 
}}

\section{Conclusions}\label{sec:conclusion}
\siweir{In this paper, we explored how to leverage a GNN model to generate pre-trained embeddings using the existing social relations among users. \io{Next}, we \io{used} the Gaussian Mixture Model to carefully extract prior knowledge contained in those pre-trained embeddings for the subsequent fine-tuning and recommendations. Our proposed \textit{Social-aware Gaussian Pre-trained} \io{(SGP)} model can significantly outperform competitive baselines, as demonstrated by the extensive experiments conducted on \swai{three} public datasets. Furthermore, a detailed user analysis showed that by incorporating the social relations, users who have \swr{less than 10} interactions particularly benefit from our SGP model. Moreover, we \swr{\io{showed that} our SGP model can} practically serve \textit{extreme cold-start} users with reasonable recommendations when \io{it} only knows about the friend's preferences of these newly registered users. Finally, we \io{used} an ablation study to examine the effect of social relations on our proposed model and a hyperparameter analysis to study the effects of different cut-offs and embedding dimensions, followed \swr{by the visualisation of the generated embeddings} \swr{to further illustrate} how our proposed model could benefit recommendations.} \swext{As future work, we aim to investigate how to leverage the bidirectional nature of social relations so that we can alleviate the issue of ’trust’/’trustedby’ relations, as mentioned in Section~\ref{sec:ablation}.} \blue{
In addition, we aim to leverage more effective fine-tuning methods for SGP, such as the contrastive graph learning method~\citep{yang2023generative}, instead of the plain training method.}

\bibliographystyle{cas-model2-names}
\bibliography{reference}

\end{document}